\documentclass[useAMS,usenatbib]{mn2e}
\usepackage{graphicx}
\title[SDSS 143027.66-005614.8 ]{Mergers and interactions in SDSS type 2 quasars at $z\sim$0.3-0.4.  SDSS J143027.66-005614.8: a case study. \thanks{Based on data from: the Sloan Digital Sky Survey,  the
European Southern Observatory (Paranal, Chile) (programme 087.B-0034) and the Hubble Legacy Science Archive (programme 10880).}}
\author[Villar-Mart\'\i n et al.]{M. Villar-Mart\'\i n$^{1,2}$, A. Cabrera Lavers$^3$, P. Bessiere$^4$, C. Tadhunter$^4$ 
\newauthor M. Rose$^4$, C. de Breuck$^5$ \\
$^{1}$Instituto de Astrof\'\i sica de Andaluc\'\i a (CSIC), Glorieta de la Astronom\'\i a s/n, 18008 Granada, Spain. villarmm@cab.inta-csic.es \\
$^2$Centro de Astrobiolog\'\i a (INTA-CSIC), Carretera de Ajalvir, km 4, 28850 Torrej\'on de Ardoz, Madrid, Spain \\
$^3$Instituto de Astrof\'\i sica de Canarias, C) V\'\i a L\'actea s/n, La Laguna, Tenerife (Spain) \\
$^4$Dept. of Physics \& Astronomy, Hicks Building, Hounsfield Road, Sheffield, S3 7RH, UK \\
$^4$European Southern Observatory, Karl Schwarschild Str. 2, D85748, Garching bei M\"unchen, Gemany}

\begin{document}

\date{} 

\pagerange{\pageref{firstpage}--\pageref{lastpage}} \pubyear{2002} 

\maketitle

\begin{abstract}

We present a compilation of HST images of 58 luminous SDSS type 2 AGNs at 0.3$\la$z$\la$0.4.
42 of them are type 2 quasars, which are a good representation of all optically selected SDSS type 2 quasars in this $z$ range. We find that the majority of the host galaxies are ellipticals (30/42 or 71\%). This is consistent with studies of radio loud and radio quiet type 1 quasars which show that their host galaxies are in general ellipticals.

A significant fraction of type 2 quasars ($\ge$25/42 or $\ge$59\%) show clear signatures of morphological disturbance which are in most cases identified with merger/interaction processes. We discuss this in the context of
 related works on type 2 quasars and powerful radio galaxies.

We study in detail the particular case of the radio quiet type 2 quasar SDSS J143027.66-005614.8 at $z=$0.318 based on  VLT, HST and SDSS imaging and spectroscopic data. The system shows highly complex morphology, similar to that found in many ULIRGs, which suggests that it is in the late pre-coalescence stage of a major galaxy merger. The optical continuum spectrum is dominated by a young stellar population of age $<$80 Myr, probably formed as a consequence of a merger induced starburst.

Ionized gas is confirmed up to a maximum total extension of $r\sim$13 kpc from the quasar, although gas is also possibly detected at up to $r\sim$32 kpc. The ionizing mechanism,
AGN vs. stellar photoionization, varies depending on the spatial location. There is a nuclear ($r\le$few kpc) ionized outflow, which is blueshifted by $\sim$520 km s$^{-1}$ relative to the
systemic redshift and has FWHM$\sim$1600 km s$^{-1}$. Several $\times$ 10$^5$ M$_{\odot}$ at most are expected to participate in the outflow.

We discuss the global properties of the object in the context of theoretical and observational studies  of galaxy mergers/interactions and their role in the triggering of the nuclear and star formation activities in the most luminous active galaxies.

\end{abstract}


\label{firstpage}

\section{Introduction}

According to the standard unification model of active galaxies (AGNs)  certain classes of type 1 and type 2 AGNs are the same entities, but have different orientations relative to the observer's line of sight (e.g. Seyfert 1s and 2s, Antonucci \citeyear{ant93}). An obscuring structure blocks the view to the inner nuclear region for certain orientations.  The existence of a high luminosity family of type 2 (i.e. obscured) quasars was predicted a long time ago, based on the AGN unification model. Indeed, narrow line radio galaxies have been known for decades, but the radio quiet counterparts, known as type 2 quasars, 
have been discovered in large quantities at different wavelengths only in the last decade:   X-ray (e.g., Szokoly
et al. \citeyear{szo04}), infrared (e.g. Mart\'\i nez-Sansigre et al. \citeyear{san05},  Stern et al. \citeyear{ste05}) and optical wavelengths 
(Reyes et al. \citeyear{rey08}, Zakamska et al. \citeyear{zak03}).   These  last authors  have identified nearly 1000 type 2 quasars  at redshift 0.2$\la z\la$0.8 in the Sloan Digital Sky Survey (SDSS, York et al. \citeyear{york00}) based on their optical emission line properties:  narrow H$\beta$ FWHM,  high ionization emission lines characteristic of type 2 AGNs and narrow  line  luminosities typical of type 1 quasars. 

Since their discovery an intensive follow up has been carried out by different research groups covering a wide spectral range  to characterize the general properties of type 2 quasars and test the AGN unification model.  Type 2 quasars are, moreover, excellent scenarios to investigate fundamental issues related to quasars and galaxy evolution in general, such as  the role of mergers/interactions to
 trigger the nuclear activity, feedback and star formation in the most powerful active galaxies.  The fortuitous occultation
of the active nucleus acts like a natural `coronograph", allowing a detailed investigation of the properties of the
surrounding medium. This is complex in type 1 quasars due to the dominant contribution of the quasar point spread function (PSF).

Based on diverse studies of several samples it can be said that:  the host galaxies are  often ellipticals with strong deviations from de Vaucoleurs profiles and frequent signatures of mergers/interactions (Zakamska et al. \citeyear{zak06}, Villar-Mart\'\i n et al. \citeyear{vil11a}). In particular,
\cite{bess12} found such evidence in 75\% of a complete sample of SDSS type 2 quasars with mean redshift $z\sim$0.35. Very intense star formation activity
 is also frequently found  (e.g. Hiner et al. \citeyear{hin09}, Zakamska.  et al. \citeyear{zak08}, Lacy et al. \citeyear{lacy07}).
 Nuclear ionized gas outflows are an ubiquitous phenomenon (Villar-Mart\'\i n et al. \citeyear{vil11b}, Greene et al. \citeyear{green11}).
 The optical continuum is often polarized, revealing the presence of an obscured luminous continuum source (Zakamska et al. \citeyear{zak05}), further reinforcing that these truly are obscured type 1 AGNs.   
  They have  large bolometric luminosities ($L_{bol}\ga 10^{45}$ erg s$^{-1}$), typical of type 1 quasars.
  They show a wide range of X-ray luminosities and obscuring column densities (Ptak  et al. \citeyear{ptak06}, Zakamska  et al. \citeyear{zak08}). 
     The detection rate
 at 1.4 GHz   in the FIRST survey is $\sim$49\%.  15\%$\pm$5\% qualify as radio loud  (Zakamska et al. \citeyear{zak04},  Lal \& Ho \citeyear{lal10}).

 Detailed studies of individual sources contribute to the understanding and characterization of the nature an phenomenology associated with type 2 quasars.
We present here results  based on imaging and spectroscopic data on the  radio quiet (Lal \& Ho 2010) type 2 quasar SDSS J143027.66-005614.8 
(SDSS J1430-00 hereafter) at $z=$0.318. This object offers an excellent opportunity to investigate in detail the triggering of the quasar activity, and the impact that this activity has on the ISM in the host galaxy.
The [OIII] luminosity as measured from the SDSS spectrum is  10$^{8.44}$ L$_{\odot}$. \cite{rey08} place a lower limit on the [OIII]5007  luminosity of $\sim$10$^{8.3}$ to select the quasars in their catalogue, ensuring that the bolometric luminosities are above the classical Seyfert/quasar separation of 10$^{45}$ erg s$^{-1}$.

  We also present a compilation of unpublished  HLA (Hubble Legacy Archive) WFPC2 and ACS/WFC images of SDSS type 2 quasars at 0.3$\la z \la$0.4.  By means of the morphological classification, we will investigate the nature of the host galaxies and the evidence for mergers and interactions.  This work complements recent studies about the incidence of mergers/interactions  and the role they play in triggering  the nuclear activity in the most powerful active galaxies  (Bessiere et al. \citeyear{bess12}, Ramos-Almeida et al. \citeyear{ram11}).

 We assume
$\Omega_{\Lambda}$=0.7, $\Omega_M$=0.3, H$_0$=71 km s$^{-1}$ Mpc$^{-1}$. At $z=$0.318,
 1" arcsec = 4.60 kpc.

\section{Observations and data reduction for SDSS J1430-00.}

The VLT long slit spectroscopic and imaging data were obtained with the Focal Reducer and Low Dispersion Spectrograph (FORS2)
installed on UT1 (Appenzeller et al. \citeyear{appen98}).
The observations were performed on 2011 April 27th and 28th for the ESO programme 087.B-0034(A).

The spectra were obtained with the grism GRIS\_300I+11 and the order sorter filter OG590+32. 
A 1.3" wide slit was used  at a position angle (PA) 9.8$\degr$ N to E. The final useful spectral range was $\sim$5900-10600  \AA\, corresponding to $\sim$4475-8040 \AA\  rest frame for $z$=0.318. The pixel scales were 
0.25$\arcsec$ pix$^{-1}$ and 3.18 \AA\ pix$^{-1}$ in the spatial and spectral directions respectively. The total exposure time was 3$\times$700=2100 seconds. The seeing size 
during the spectroscopic observations was FWHM=0.73$\pm$0.05". Point sources, therefore, did not fill the  slit. The impact of this  on the kinematic measurements will be discussed
when appropriate. The spectral resolution as measured from the arc emission lines  was    FWHM=13.6$\pm$0.2 \AA. 

 The spectra were reduced and analysed following standard procedures with  IRAF and STARLINK packages. The spectra were debiased, flat fielded, wavelength calibrated, combined with simultaneous cosmic ray removal, background subtracted, flux calibrated and corrected for Galactic extinction (E(B-V)=0.047). Geometric distortion was found to be negligible.   The flux calibration, based on the standard star LTT3864,
produced an accuracy    of $\sim$5\% over the entire spectral range. 

\begin{figure*}
\includegraphics[width=18cm]{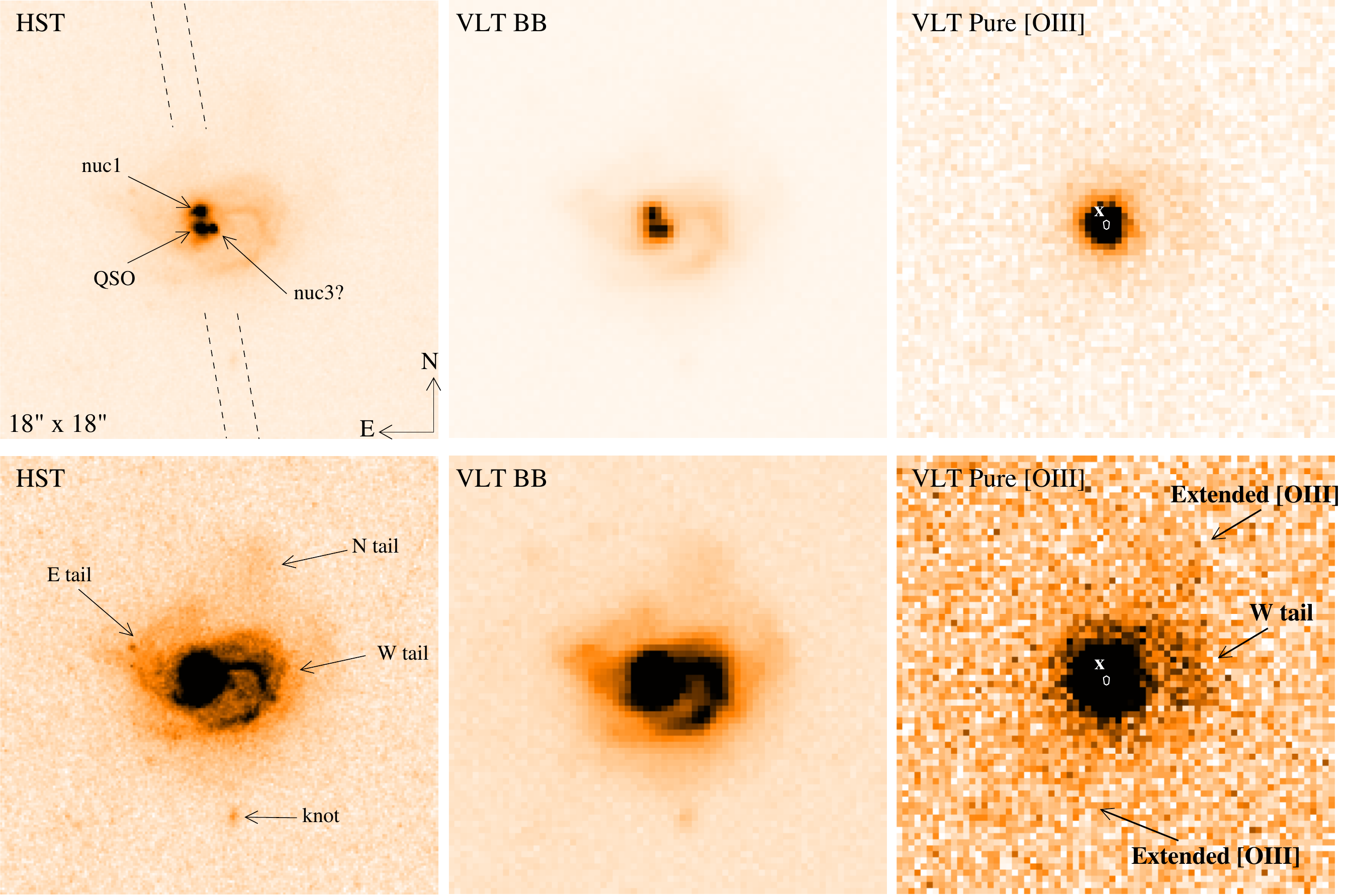}
\caption{All panels show the same 18"$\times$18" (83 kpc $\times$ 83 kpc) field centered on SDSS J1430-00.  Each pair of top and bottom panels correspond to the same image shown with different contrasts to highlight the high (top) and low (bottom) surface brightness structures. Left: HST broad band image (F814W filter).
Middle: VLT broad band image (V\_high filter). Right: VLT pure (i.e. continuum subtracted) [OIII]$\lambda$5007 narrow band image (see text). The main morphological features discussed in the text are indicated.``x" and ``o" in the [OIII]  image indicate the location of $nuc1$ and the QSO continuum centroid. The approximate slit location of the FORS2 spectrum is also shown in the the first panel.} 
\end{figure*}

A broad band image was created by combining 3$\times$200 = 600 sec exposures  obtained with the V\_{HIGH}+114 FORS2 VLT filter, which has central $\lambda$ 5550 \AA\ and FWHM  1232 \AA. The narrow band image, which is centered at the redshifted [OIII]$\lambda$5007 line ([OIII] hereafter), was obtained with the FORS2 interference filter  H\_Alpha/2500+60, which has central $\lambda$  6604 \AA\ and FWHM 64 \AA. Three exposures were also obtained, 
completing  a total exposure time of 3$\times$300=900 seconds.
For both the broad and narrow band images a dithering pattern of  three positions was applied.
The individual images were debiased, flatfielded and spatially registered before being combined.  

A WFPC2 HST image  obtained with the F814W filter was retrieved from the Hubble Legacy Archive.  This has central $\lambda$ 7905 \AA\ and FWHM 1539 \AA.  The only processing we applied to the HST image was cosmic ray removal.  It has a 0.1"  pixel scale. This image was obtained
on  2007  April 5th for the HST programme with identification 10880 (Principal Investigagor, PI: Henrique Schmitt).

We  also used the SDSS spectrum, which covers the 3800-9180 \AA\ spectral range.  The spectral resolution at the redshifted [OIIII] wavelength is $R=\frac{\lambda}{\Delta\lambda}\sim$1850. It was corrected for galactic extinction.

We show in Fig. 1 the HST (left panels) and VLT  broad band (middle) images.  Each pair of top and bottom panels  correspond  to the same image, shown with two different contrasts to highlight the low (top) and high (bottom) surface brightness features. The HST image, which covers the rest frame range $\sim$5415-6583 \AA\  is contaminated by several strong emission lines such as [OI]$\lambda$6300, H$\alpha$, [NII]$\lambda\lambda$6548,6583 ([NII] hereafter). The broad band VLT image (3744-4679 \AA\ rest frame)
only includes  several faint emission lines such as [NeIII]$\lambda$3869, H$\delta$, H$\gamma$, [OIII]$\lambda$4363 and HeII$\lambda$4686, which are not likely to make a substantial relative contribution to the total flux within the filter. It  is  thus expected that continuum emission dominates at most, if not all spatial locations. 

A pure [OIII]$\lambda$5007  narrow band image (right panels, Fig.~1) of SDSS J1430-00 was created by subtracting the VLT narrow  and broad band images. These  were first registered spatially and the fluxes scaled so that stars in the field leave
minimum residuals when subtracted.

\section{Results}

\subsection{A merging system.}

The system shows clear signatures of a merger (Fig.~1), such as  tidal tails and at least two nuclei separated by $\sim$0.75" or 3.5 kpc. As mentioned in the introduction, evidence for mergers and interactions is frequent in optically selected type 2 quasars.

 Four tidal tails are identified extending in different directions (N, S, E and W tails in Figs.~1 and 3). The longest tail, which spreads to the South for at least $\sim$22" or 100 kpc from the QSO, is most clearly appreciated in the broad band VLT image (Fig.~3).  This image does not contain strong emission lines, so the tail emits continuum. Its non detection in the pure [OIII] image indicates that it is not a strong line emitter.

\begin{figure}
\includegraphics[width=8cm]{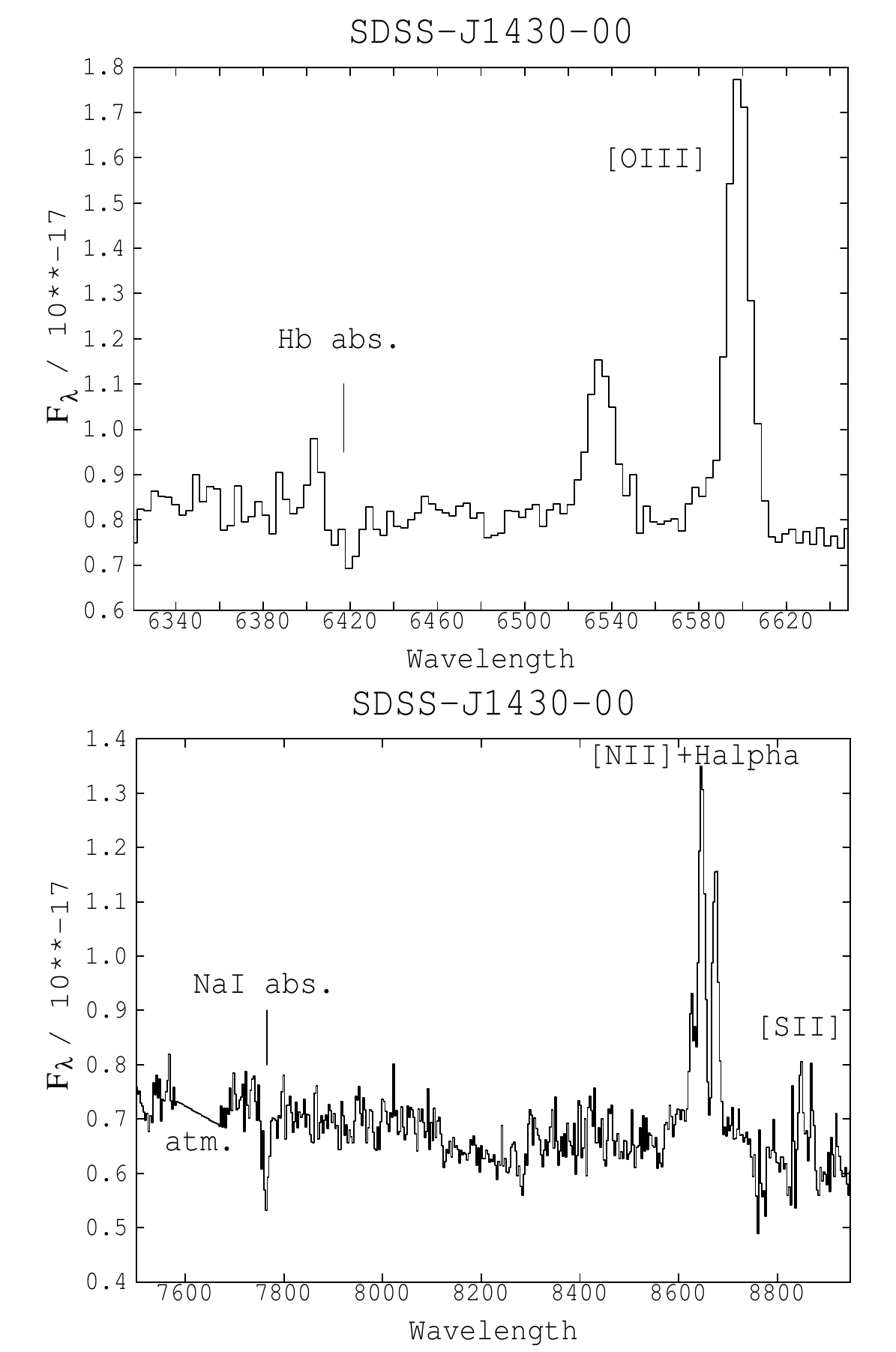}
\caption{VLT-FORS2 spectrum of the brightest continuum pixel of $nuc1$. Although the emission lines are due to contamination by the QSO,
both H$\beta$ and NaI appear in absorption, confirming that $nuc1$ is associated with the quasar.``atm" indicates residuals of an atmospheric band.
The monochromatic flux $F_\lambda$ is given in units of 10$^{-17}$ erg s$^{-1}$ cm$^{-2}$ \AA$^{-1}$ and the wavelength in \AA.} 
\end{figure}

Three compact sources
are identified, marked as $nuc1$, QSO (the quasar) and $nuc3?$ in the HST image.  $nuc3$  has been marked with a ``?" because its $z$ is unknown.
It is also possible that a dust line is responsible for this apparent detachment between $nuc3$ and the QSO.

As we will see in $\delta$3.3, no line emission can be confirmed associated with $nuc1$, neither from the images, nor from the spectrum,
so its $z$ cannot be determined this way.
We have extracted an one dimensional spectrum of the spatial pixel where the $nuc1$  continuum is brightest  (Fig.~2.) 
Although the QSO and $nuc1$ are spatially resolved in the VLT broad band image (Fig.~1), and also in the VLT spectrum at continuum wavelengths (Fig.~7), the bright QSO line emission  spreads over $nuc1$ and contaminates it  ($\delta$3.2). However, both H$\beta$ and NaI are detected in absorption, implying  $z$=0.3176$\pm$0.0006 similar to that of the QSO. $nuc1$ is thus, a companion nucleus. The projected separation relative to the quasar nucleus is $\sim$0.75" or 3.5 kpc.

  A large low surface brightness  halo, which presents  rich morphological substructure, is detected extending for  $\sim$14" (64 kpc)  N-S and 15" (69 kpc)  E-W in the broad band VLT image.  The knot identified in both the HST and the VLT images is roughly aligned with the QSO-$nuc1$ line. Compact knots of star formation have been found associated with some type 2 quasars (Villar-Mart\'\i n et al. \citeyear{vil11a}), however it is not certain whether in this case
it belongs to the same system. The spectrum shows a very faint continuum source with no confirmed absorption or emission lines.

  From the morphology of the features, this is  the pre-coalescence stage of a
merger in which the nuclei have not merged yet.

\begin{figure}
\includegraphics[width=8cm]{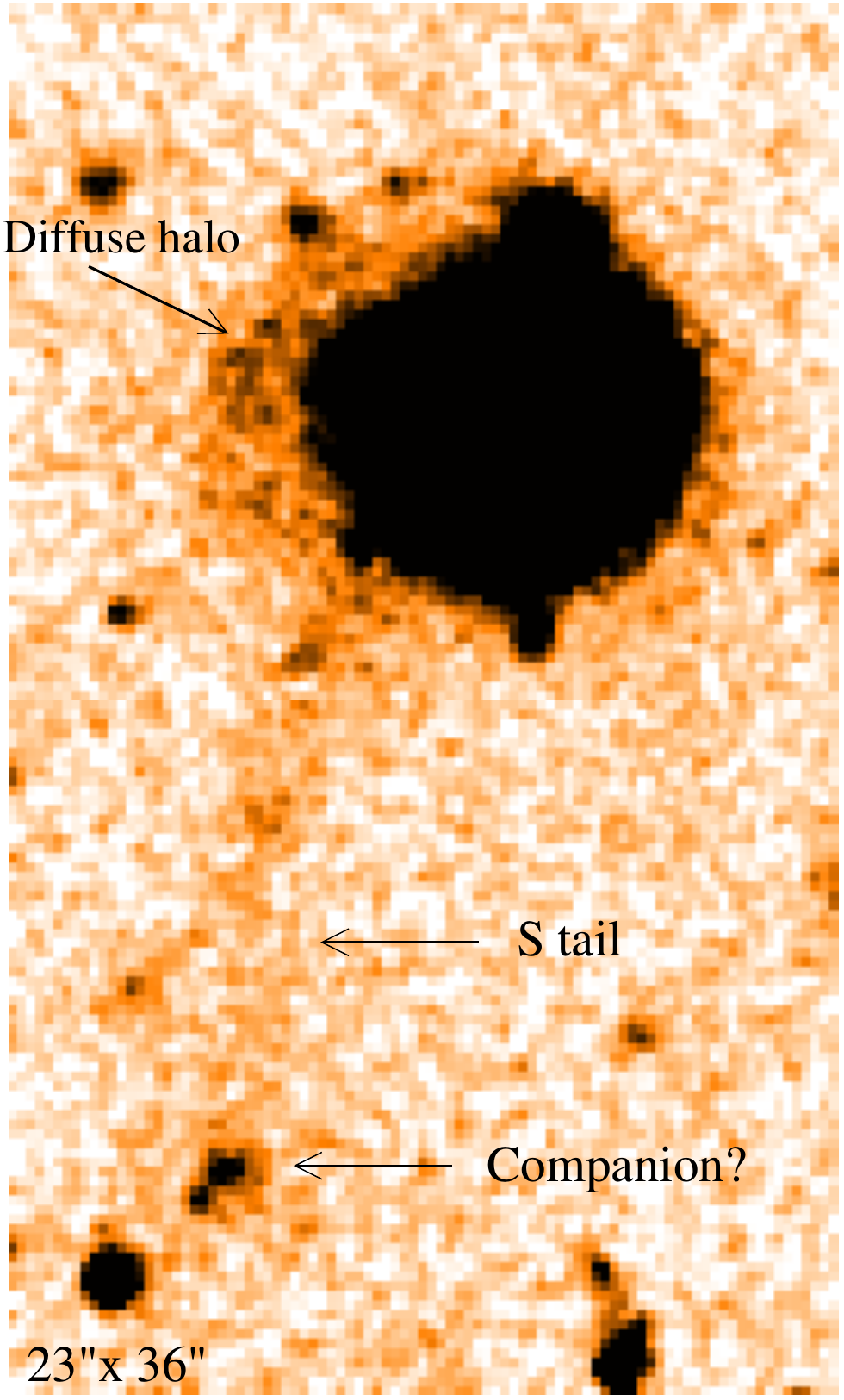}
\caption{SDSS J1430-00 VLT broad band image smoothed with a median filter of 3$\times$3 window size, to highlight the Southern tail and the diffuse low surface brightness halo. The  effective seeing is 0.85".} 
\end{figure}

\subsection{The stellar populations}

We have  determined the age of the stellar populations in SDSS J1430-00 by means of fitting the spectral energy distribution of the SDSS spectrum (thus, this information refers to an aperture of 3" ($\sim$14 kpc) diameter centered on the quasar).
The SDSS spectrum covers a wider $\lambda$ range than the VLT spectrum, down to the blue below the 4000 \AA\ break.

\begin{figure*}
\includegraphics[width=17cm]{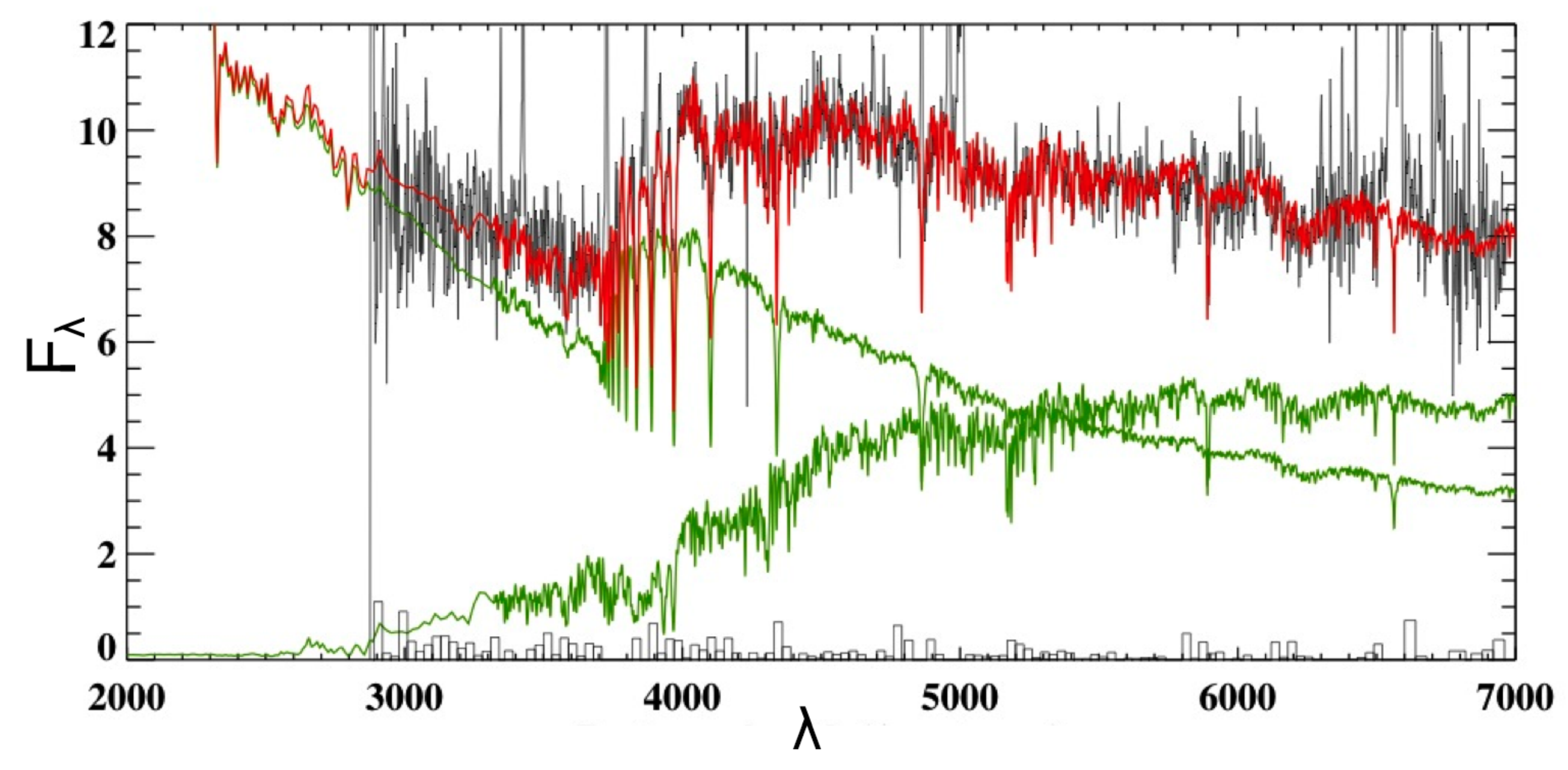}
\caption{The stellar populations. Typical output from CONFIT showing the original SDSS spectrum (black), the composite model fit (red) and the model templates of the OSP and YSP at the level of their relative contribution to the total flux (green). In this case, we have an OSP of 12.5 Gyrs and a YSP of 0.04 Gyrs with E(B-V)=0.1.  The residuals (data - model fit) are shown in the lower part of the figure (black). $F_{\lambda}$ in units of 10$^{-17}$ erg cm$^{-2}$ s$^{-1}$ \AA$^{-1}$ and
$\lambda$ in \AA.} 
\end{figure*}

For this purpose we have used CONFIT, which is a purpose written IDL based code. For full details of its functions and limitations, refer to 
\cite{rob00} and \citet{holt07}. To summarise, CONFIT uses a minimum $\chi^2$ technique in order to find the best fit to spectra, using up to three components. These components are an old stellar population (OSP), which is usually assumed to be 12.5 Gyrs old, a young stellar population (YSP)  that can vary between 0.001-5.0 Gyrs
(model templates taken from Bruzual \& Charlot \citeyear{bruz03}). Reddening of $0.0<E(B-V)<1.6$ is also applied to the YSP in increments. If necessary, a power-law component of the form $F_{\lambda} \propto \lambda^{\alpha}$ can also be fit, which can represent an AGN/scattered light component, or a very young stellar population. 

Prior to the modeling it was important to gauge the level of the nebular continuum component, since this component is known to make a significant contribution
to near-UV continua of narrow line AGN with high equivalent width emission lines (Dickson et al. \citeyear{dick95}). In the case of SDSS J1430-01 we find that, based on the
absorption corrected H$\beta$ flux, the nebular continuum is negligible, contributing $<$5\% of the continuum below 3500\AA. Therefore no correction for the nebular continuum was necessary.
In order to model the continuum spectrum, 102 continuum bins of 30 {\AA} width were used, chosen to avoid emission lines. The full wavelength range of the SDSS spectra ($\sim3000-7000$ {\AA} in the rest frame) was used so that we could obtain the best possible modeling fit. The spectral bin $5050-5080$ {\AA} was
selected for normalisation. CONFIT scales the flux from the model components so that the total flux incorporated in the model is always less that 125\% of the observed flux. A relative flux calibration error of $\pm5\%$ is assumed. The code then calculates the minimum $\chi^2$ for each combination of components using different relative fluxes between the components.

We first modeled the spectra using a combination of an OSP and a YSP (Fig.~4). 
It is found that two sets of models produce an adequate fit to the
SDSS spectrum ($\chi^2_{red} < 1$ in Fig.~5). The first corresponds to a combination of an OSP of age 12.5 Gyrs and a YSP of age 4 -- 7 Myr with with moderate reddening $0.2 < E(B-V) < 0.5$. The second comprises an OSP of 12.5 Gyrs, plus a  YSP of age 30 -- 70 Myr with reddening $0.0 < E(B-V) < 0.4$. In the case of these two component fits, the older YSP age (30 -- 70 Myr) is favoured because it provides a better fit to detailed features, such as CaII K$\lambda3968$ and the higher order Balmer lines, which are sensitive to stellar age. Note that, with this model combination, older YSP ages $>$80~Myr are ruled out.

We also investigated whether successful models could be obtained by including a power-law component which would account for the possible presence of scattered
light from the hidden quasar. In this case models including older YSP ages (up to $\sim$1~Gyr) can provide good fits to the spectrum. However, for models with YSP ages greater than 60 Myrs, the contribution of the power law component to the total flux becomes large (30 -- 50\% at 4860 \AA). If the scattered quasar does indeed contribute this proportion of the flux, we would expect to see evidence for a broad H$\beta$ feature in the model-subtracted SDSS spectrum. Since no such feature is detected, we conclude that YSP ages $\le$60~Myr
-- consistent with the results of the 2 component fits described above -- are favoured by the fits that include a power-law component; for such ages, the power-law contributes $<$20\% of the flux at 4860\AA, consistent with the
non-detection of a scattered BLR component to H$\beta$.

Overall, we conclude that, in addition to an OSP of age $\sim$12.5~Gyr, a YSP of age $<$80~Myr is also present. This latter component is likely to have formed as a consequence of a starburst induced by the merger.

  \begin{figure}
\includegraphics[width=8cm]{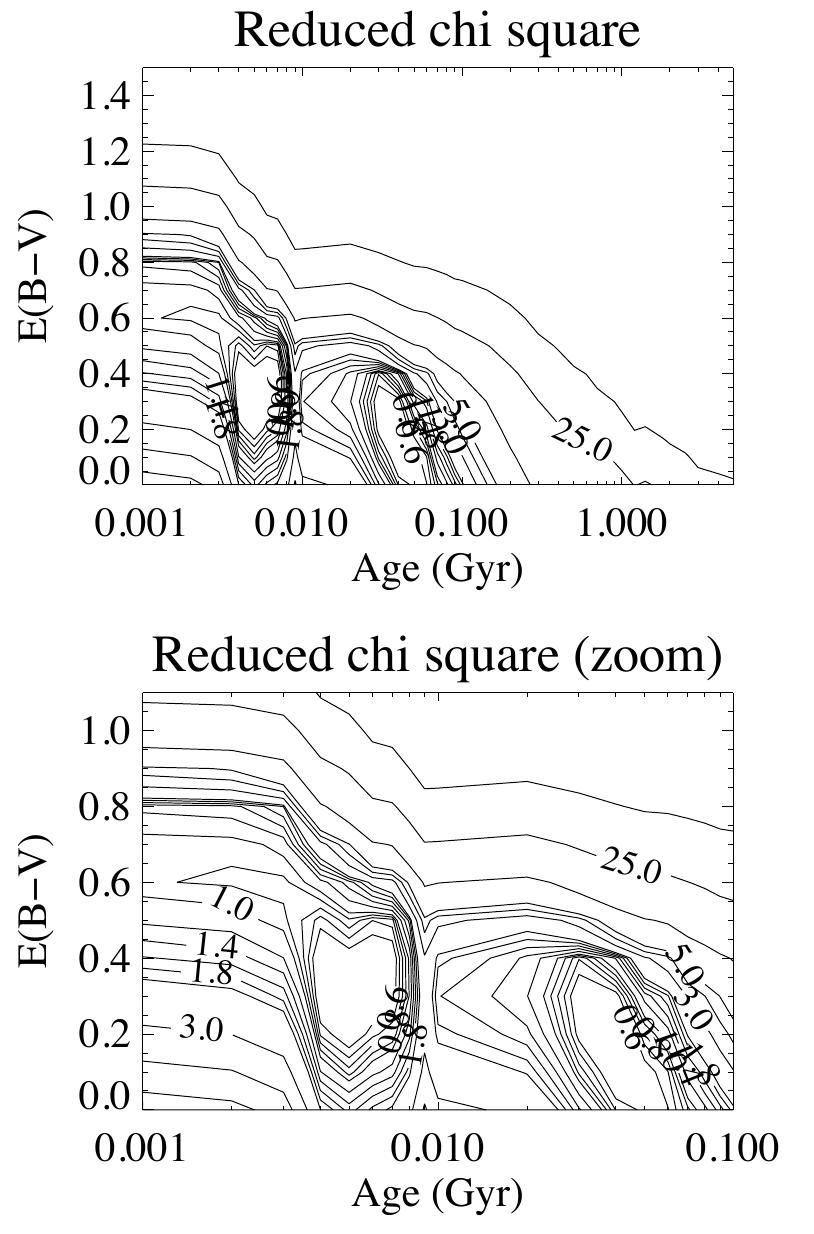}
\caption{A contour plot showing the reduced $\chi^2$ for the different model fits showing the age of the YSP against reddening E(B-V) of the YSP. The top panel shows the results for all possible combinations, whilst the lower panel zooms in on the most likely solutions. These are the model solutions including only an OSP and a YSP. Two minima are found: the first  corresponds to a YSP of age$\sim$4-7 Myrs  and
 E(B-V)$\sim$[0.2-0.5]. The second minimum corresponds to a YSP of age$\sim$30-70 Myrs 
and  E(B-V)$\sim$[0.0-0.4]. } 
\end{figure}

\subsection{The ionized gas: spatial distribution, ionization mechanism and physical conditions.}

The [OIII]$\lambda$5007 continuum subtracted image (Fig.~1, right panels)  shows that the  line emission  is dominated by a compact region associated with the quasar, which spreads over $nuc1$.   
Extended [OIII]  is  also hinted, although very faint,  overlapping with the W tidal tail and at several locations of the large diffuse halo mentioned above. Unlike the continuum halo, this  halo line emission  is elongated along PA 23 N to W, where it reaches a maximum extension of $\sim$13" or $\sim$60 kpc. It overlaps to the North with the N tail and extends also at the opposite side of the QSO nucleus. This elongated morphology at both sides of the QSO suggests that this gas is  within the AGN ionization cone. On the contrary, the location and morphology of the  line emission from  the W tail rather
indicates that this gas is excited locally by young stars.

 We show in Fig. 6 the 2D VLT long slit spectra of H$\beta$, [OIII]$\lambda\lambda$4949,5007 and the [NII]+H$\alpha$ lines
 along PA 9.8 N to E.   Line emission is detected up to a maximum extension of  $\sim$5" (23 kpc): 2.1" to the N and 2.9"  to the  S of the quasar continuum centroid.  The extended lines along this PA are well within the size of the diffuse halo (Fig. ~1). The lines
 are noticeably brighter towards the South. 
 
 The spatial profiles of [OIII]$\lambda$5007 (blue), the [NII]+H$\alpha$ complex (red) and the continuum (black) along the spectroscopic slit are shown in Fig.~7.  The emission line profiles have been continuum subtracted, using the continuum extracted from an adjacent  window of similar
spectral  width. The  locations of the
 quasar continuum centroid  and $nuc1$ are indicated with the spatial axis zero    and the vertical dotted line respectively.
 Continuum is detected for a total extension of at least  9"or 41 kpc along the slit, similar to the extension  of the low surface brightness  halo
 along the same PA seen in the images.  
   As already concluded from the images, no line emission from $nuc1$ is confirmed.   

  \begin{figure*}
\includegraphics[width=18cm]{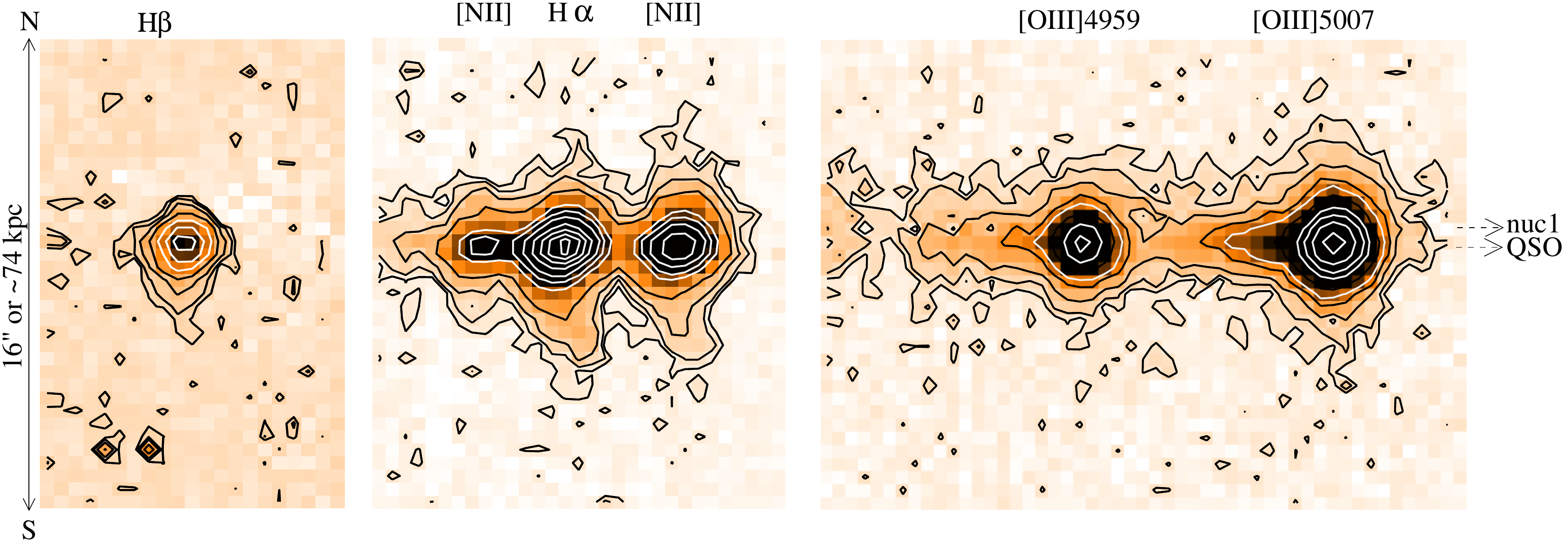}
\caption{2-dim spectrum of the main optical emission lines. The adjacent continuum has been subtracted in all cases. The lines are clearly extended.
The location of the quasar continuum centroid (QSO) and the second nucleus are indicated. Broad blue wings are appreciated in the [OIII] and H$\alpha$ lines. The contours in the [NII]-H$\alpha$ 2-dim spectrum have  levels 60, 50, 40, 30, 20, 14, 10, 4, 2 and 1 times brighter than the most external, faintest contour. These values are 200, 160, 100, 60, 50, 20, 12, 8, 4, 2, 1 for
the [OIII] contours and 28, 16, 13.5, 8, 4, 2 and  1 for H$\beta$.} 
\end{figure*}

 The line emission  is dominated by a spatially compact component, associated with
 the quasar nucleus. It has a spatial FWHM of 0.75$\pm$0.05$\arcsec$, similar to  the seeing size 0.73$\pm$0.05; thus, it 
 is unresolved. In addition, extended low surface brightness emission is detected up to a maximum extension
 of 3$\arcsec$ or $\sim$14 kpc from the quasar continuum centroid towards the South. 
  
  There is a bump of both [NII]+H$\alpha$ and continuum emission at $\sim$1.5-2" of the QSO (Fig.~7), coinciding with the W tidal tail. This feature wraps around   the QSO clockwise towards the South and part of it is within the slit (Fig.~1, top left). Such excess of both line and continuum emission supports, as suggested by
  the [OIII] image, that the gas in the Western tail is preferentially ionized locally by young stars. This implies that star formation occurs   at distances $\sim$15 kpc from the quasar location, associated with  a tidal tail.

The line ratios add further information on the ionization mechanism of the gas at different spatial locations.   Line ratios relative to H$\beta$ are shown in Table 1 as measured with the SDSS spectrum. The H$\beta$ and H$\alpha$ fluxes have been corrected for underlying stellar absorption using the best continuum model fits discussed in $\delta$3.2.
According to this range of models, the H$\beta$ and H$\alpha$ fluxes are absorbed by 28$\pm$5\% and 10$\pm$2\% respectively.
 The Balmer decrement H$\alpha$/H$\beta$ was used to correct
for reddening (Osterbrock \citeyear{ost89}). The detection of [NeV] and HeII implies that the nuclear gas is photoionized by the quasar,  as naturally expected based on the selection criteria
applied by Zakamska et al. (2003) to build the SDSS type 2 quasar sample.

\begin{table}
\centering
\begin{tabular}{llll}
\hline
Line/H$\beta$ &  Observed & Dered.  \\ \hline
~[NeV]$\lambda$3426 & 0.61$\pm$0.06 &  1.0$\pm$0.1 \\
 ~[OII]    &  1.11$\pm$0.07  &   1.8$\pm$0.1  \\
 ~[NeIII]$\lambda$3869  & 0.56$\pm$0.05 &    0.9$\pm$0.1 \\
 ~[OIII]$\lambda$4363   & $\le$0.06 &  $\le$0.08   \\
~HeII  &  0.28$\pm$0.03  &   0.30$\pm$0.05 \\
 ~[OIII]$\lambda$5007  & 4.3$\pm$0.1 &  4.1$\pm$0.1   \\
  ~[OI]$\lambda$6300 & 0.18$\pm$0.03 &  0.12$\pm$0.03 \\ 
 ~H$\alpha$ & 4.67$\pm$0.08 &  3.01$\pm$0.07\\
 ~[NII]$\lambda$6583 & 2.6$\pm$0.1 & 1.68$\pm$0.08  \\ 
 ~[SII]$\lambda\lambda$6716,6731 & 1.58$\pm$0.09 &   1.02$\pm$0.06\\ 
~T$_4$ & $\la$1.30 &  $\la$1.60 \\   \hline
~F(H$\beta$)  & 8.9$\pm$0.4&   & \\ 
\hline 
\end{tabular}
\caption{Nuclear line ratios relative to H$\beta$ measured from the SDSS spectrum.  The H$\beta$ and H$\alpha$ fluxes have been corrected for underlying stellar absorption (see text). The ratios corrected for dust extinction (column [3])
are based on the reddening derived from H$\alpha$/H$\beta$ (case B recombination value is 3.0). The H$\beta$ flux is given in units of 10$^{-16}$ erg s$^{-1}$ cm$^{-2}$. T$_4$ is the electronic temperature in units of 10$^{4}$ K.}
\end{table}

\begin{figure}
\includegraphics[width=8cm]{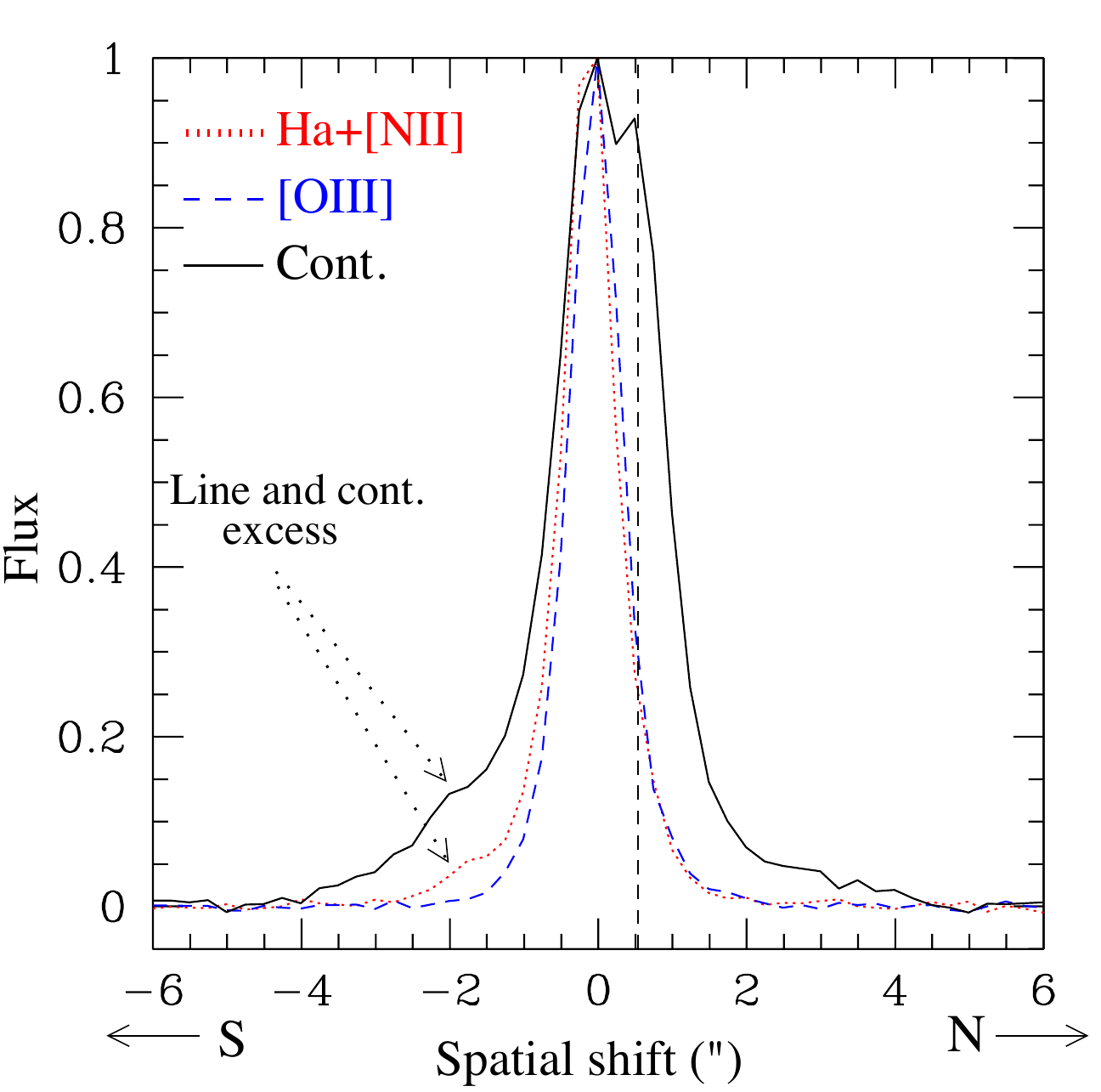}
\caption{Comparison of the spatial distribution of the emission lines (H$\alpha$+[NII] (red) and [OIII] (blue), both continuum subtracted) and the continuum (black) along the spectroscopic slit. The spatial 0 of the x axis marks the
location of the quasar continuum centroid. The vertical dotted line indicates the position  of the companion nucleus $nuc1$. In all cases the fluxes
have been normalized to the maximum value.  There is an excess of both continuum and line emission $\sim$1.5-2" to the 
South of the continuum centroid, coinciding with the location of a tidal tail. This suggests the presence of local ionization by young stars.} 
\end{figure}

We show in Fig. 8  the   [OIII]/H$\beta$ vs. [NII]/H$\alpha$  
and [OIII]/H$\beta$ vs. [SII]/H$\alpha$ diagrams, which are very efficient at discriminating whether the gas in an emission line galaxy is photoionized by young stars or an AGN  (Baldwin, Philips \& Terlevich \citeyear{bpt81}). The blue dashed lines mark the maximum starburst line derived by Kewley et al. (\citeyear{kew01}, Ke01).  The red continuous line in the top diagram marks
the starburst line derived by Kauffmann et al. (\citeyear{kauf03}, Ka03). According to these studies,   objects below the Ka03 are starburst galaxies, while objects above the Ke01 line are active galaxies. The objects in between are composite in nature. 

The location of the SDSS  spectrum (solid circle) and several apertures extracted along the FORS2 slit are shown in both diagrams with open symbols.
The QSO nuclear spectrum (open circle) was extracted from an  $\sim$1" wide aperture centered at its continuum centroid.  Both the SDSS and FORS2 QSO nuclear ratios have been corrected for underlying stellar absorption. 
The spectrum (open triangle) from the Southern extended ionized gas (EELR, extended emission line region) was extracted from a 1" wide aperture centered  1.9" to the South. To avoid as much contamination as possible from the quasar, a 0.5" wide aperture centered at the $nuc1$ centroid was used to extract the spectrum of the companion nucleus (open square).

The  quasar nuclear gas is located in the AGN area of the diagrams as expected. If the same correction for underlying stellar absorption was applied
to  $nuc1$ line ratios, they would become almost identical to those of  the quasar.
This reinforces the idea that the line emission at this location is strongly contaminated by the quasar.

 The  extended gas is  located at the edge of the transition area, suggesting that, in addition to AGN photoionization, stellar
photoionization is also  present. Both diagrams produce the same conclusion.  Given that some contamination by the quasar nuclear emission is also possible, this gas must be at least partially ionized by stars. This is consistent with the idea proposed above that  young stars are responsible for local ionization at certain spatial locations.

\begin{figure}
\includegraphics[width=8cm]{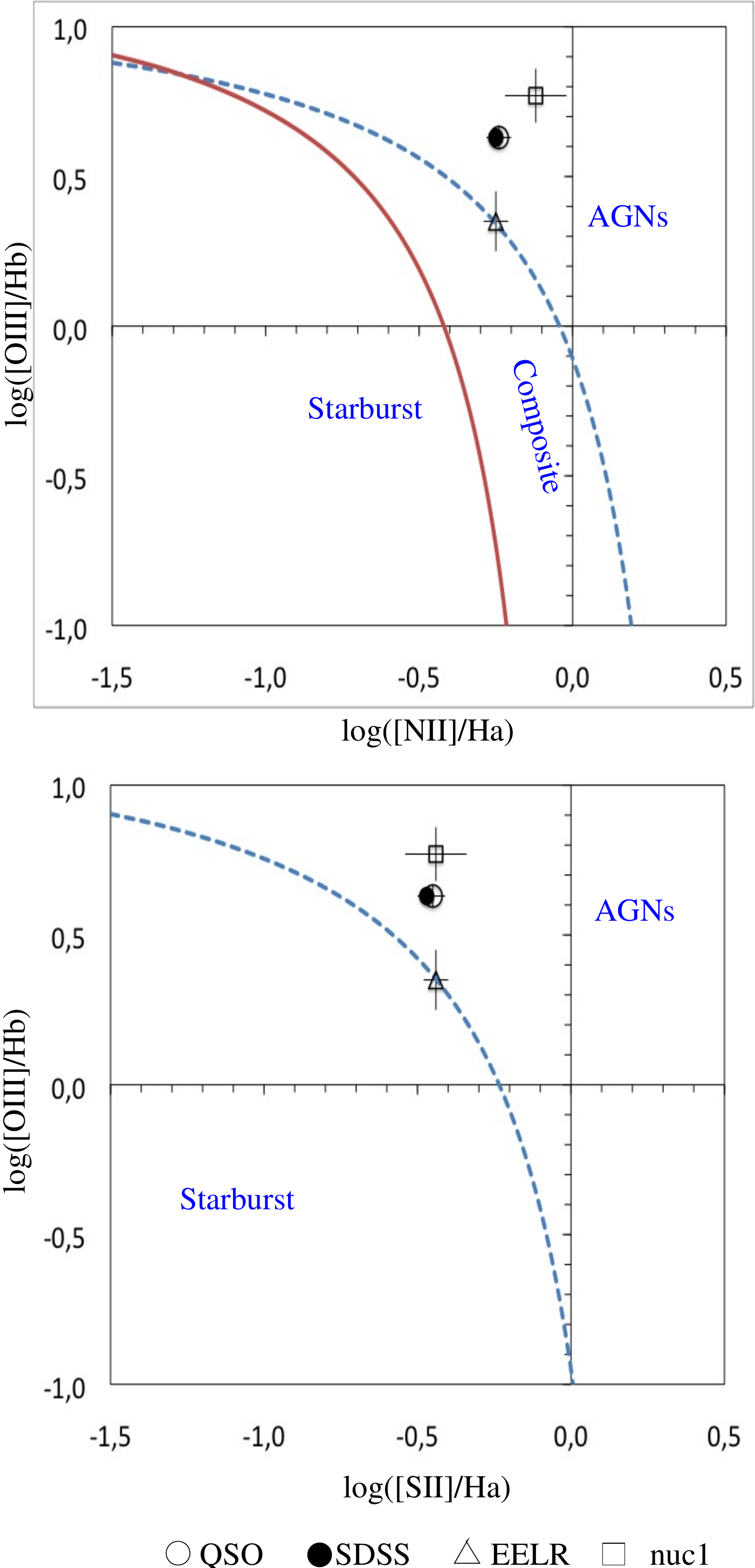}
\caption{Diagnostic diagrams to discriminate the nature of the ionization mechanism at different locations.
The blue dashed line in both diagrams marks the
maximum starburst line derived by Kewley et al. (2001).   Objects above this line are AGNs. The continuous red line on the top diagram marks
the starburst line derived by Kauffman et al. (2003). For objects below this line  the gas is preferentially photoionized by young stars. The nuclear gas 
of SDSS J1430-00 is clearly photoionized by the AGN
(also implied by the presence of strong HeII and [NeV]  emission). The extended gas (open triangle) lies in the frontier between the AGN and composite
object region. It is at least partially photoionized by stars. The companion nucleus $nuc1$ has   strong contamination 
by the QSO emission lines.} 
\end{figure}

The electron density could be measured for the nuclear gas, from [SII] 6716/6731 = 1.10$\pm$0.08, which implies a gas density  $n_e$=400$^{+200}_{-150}$ cm$^{-3}$.  
The [OIII]$\lambda$4363 line is outside the useful FORS2 spectral range. We have used the SDSS spectrum instead, where only  an upper limit could be estimated and implies an electron temperature T$_e\la$16.000 K, once reddening correction has been applied
(Table 1). The following, very faint coronal lines have been identified in the nuclear spectrum: [Fe IV]$\lambda$4903, [CaV]$\lambda$5309, [FeVII]$\lambda$5721, [FeVII]$\lambda$6087.

In summary,  the existence of ionized gas is confirmed over  a maximum total extension of $\sim$23 kpc ($r\sim$13 kpc from the quasar), although gas at distances of up to $r\sim$32 kpc is also hinted. The  ionization mechanism varies depending on the location: AGN and stellar photoionization are both present.

\subsection{The gas kinematics}

\cite{vil11b} found that nuclear outflows are an ubiquitous phenomenon in type 2 quasars. The inspection of
the nuclear emission line profiles of SDSS J1430-00 already reveals blue asymmetric profiles and the presence of broad underlying blue wings for several emission lines. According to previous results on both type 2 quasars and other active galaxy types (see VM11b for a discussion), we interpret this as  evidence for an ionized outflow in SDSS J1430-00.  In order to maximize the signal/noise for the faint wings, we have
extracted a spectrum from a 0.75''  (3.5 kpc) wide aperture centered at the quasar continuum centroid. 

The SDSS spectrum (Fig.4) shows clear stellar signatures, such as the H and K CaII lines. The first is most clearly defined.
We find that the $z$ of this absorption line and the narrow [OIII] core are shifted by 10$\pm$30 km s$^{-1}$. I.e., as argued in VM11b, it is reasonable to assume that the narrow [OIII] core is a good indicator of the systemic redshift of the galaxy. We will work
under this assumption in the following analysis.  

\begin{figure}
\includegraphics[width=9cm]{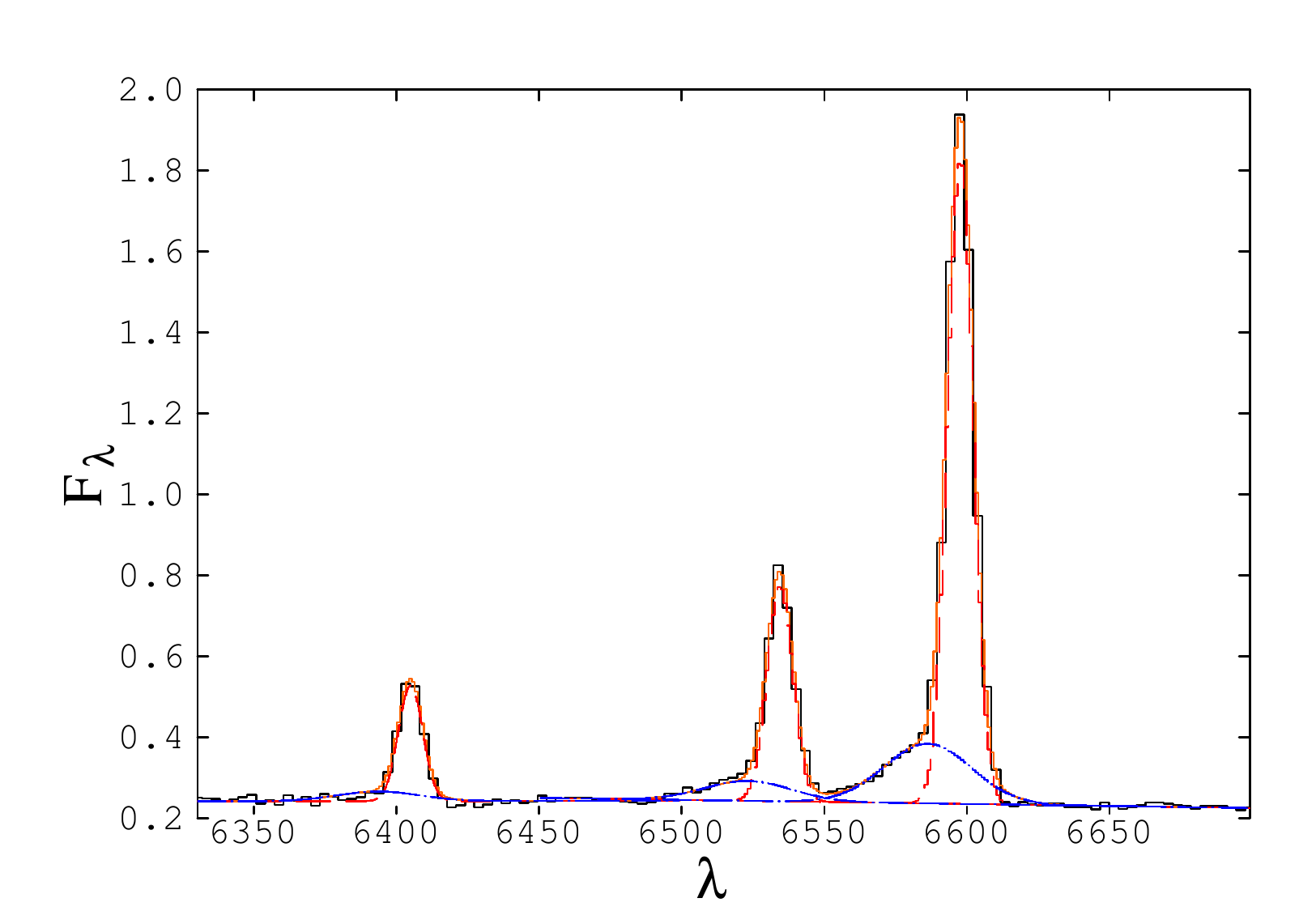}
\caption{Fit of H$\beta$ and [OIII]$\lambda\lambda$4959,5007 for SDSS J1430-00 using the VLT nuclear spectrum. The data and the fit are shown in black and orange respectively. The same colour and line style are used for each kinematic component in both emission lines: red  and blue for the most redshifted and blueshifted components respectively. $F_{\lambda}$ in units of 10$^{-17}$ erg s$^{-1}$ cm$^{-2}$ \AA$^{-1}$ and $\lambda$ in \AA.}
\includegraphics[width=9cm]{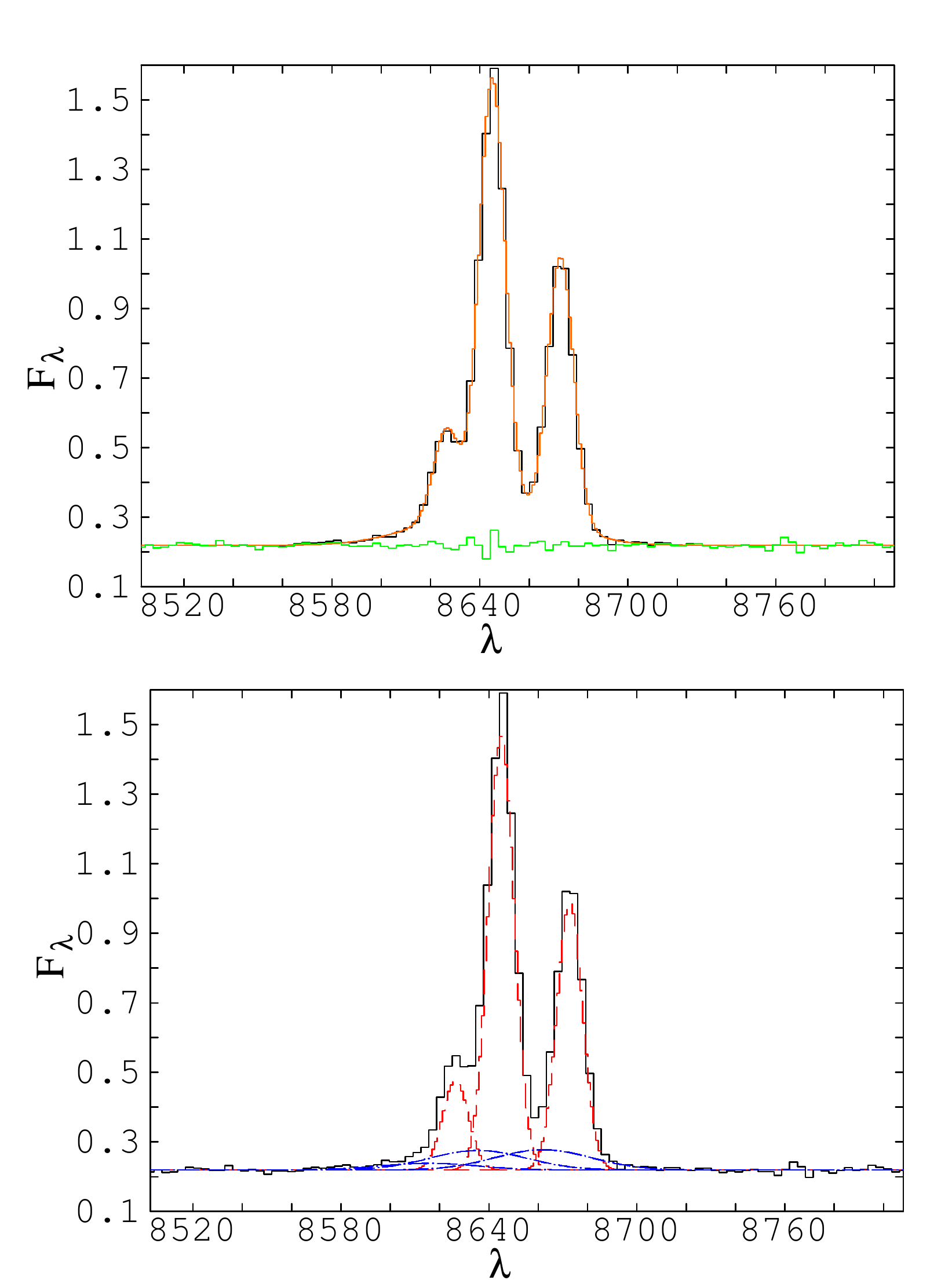}
\caption{Fit of H$\alpha$ and [NII]$\lambda\lambda$6548,6583.  Top: Data (black), fit (orange) and
residuals (green). Bottom: the individual kinematic components are shown in red and blue following the same colour code as in Fig.~9. $F_{\lambda}$  in units of 10$^{-17}$ erg s$^{-1}$ cm$^{-2}$ \AA$^{-1}$ and $\lambda$ in \AA.}
\end{figure}

We have fitted the [OIII]$\lambda\lambda$4959,5007 spectral profiles by applying the same technique as in VM11b. Two components are revealed by the fit (Fig.9). 
The spectral profile of the  lines is dominated by a
 narrow component of FWHM=10.5$\pm$0.2 \AA, clearly narrower than the IP
inferred from the arc lamp (13.6$\pm$0.2 \AA). This confirms that the
nuclear gas emission did not fill the slit. Since, as seen in  $\delta$3.3, the nuclear line emission is dominated by a 
spatially unresolved  component, for the analysis of the nuclear spectrum, 
we will assume IP=9.3$\pm$0.3 \AA\, which is the FWHM of the  seeing in pixels multiplied by the pixel scale in \AA.
So, two kinematic components are isolated in both [OIII] lines with FWHM 280$\pm$20 and 1600$\pm$200 km s$^{-1}$ respectively (quadrature correction for instrumental broadening has been applied)\footnote{Notice that the uncertainties on the IP have a very small impact  on the FWHM of the broad component, since this is dominated by kinematic rather than instrumental broadening}.  The broad
component is blueshifted relative to the narrow component by  520$\pm$60
km s$^{-1}$. 

\begin{table}
\centering
\begin{tabular}{llll}
\hline
& Narrow &  Broad  \\ \hline
   & 	& 		\\
FWHM   & 280$\pm$30	& 1600$\pm$200 		\\
   & 	& 		\\
F(H$\alpha$) & 16.4$\pm$0.3	&	2.3$\pm$0.5 \\
   & 	& 		\\
~F(H$\beta$)  &  3.04$\pm$0.06 &  $\la$0.86  & \\ 
   & 	& 		\\
$\frac{H\alpha_i}{H\alpha_{tot}}$   & 0.88$\pm$0.03	& 0.12$\pm$0.03		\\
   & 	& 		\\
$\frac{[OIII]_i}{[OIII]_{tot}}$   & 0.76$\pm$0.02	& 0.24$\pm$0.02		\\
   & 	& 		\\
$\frac{H\alpha}{H\beta}$   & 5.4$\pm$0.1 	& $\ga$2.7 		\\
   & 	& 		\\
$\frac{[OIII]}{H\beta}$   &  5.9$\pm$0.1	&  $\ga$6.6		\\
   & 	& 		\\
$\frac{[NII]}{H\alpha}$   & 0.62$\pm$0.02 	& 	1.3$\pm$0.3	\\
   & 	& 		\\
\hline 
\end{tabular}
\caption{Some measurements for the broad and the narrow kinematic components isolated in the emission lines. The line fluxes are given in units of 10$^{-16}$ erg s$^{-1}$ cm$^{-2}$ and the FWHM in km s$^{-1}$.}
\end{table}

Although  faint, the blue wing is also hinted in the H$\beta$ line. An upper limit to the flux was measured, assuming the broad
component has the same FWHM in km s$^{-1}$ as in [OIII] (Table 2).

The presence of the broad underlying wings is also evident in the [NII]+H$\alpha$ complex, which cannot be successfully fitted with single Gaussians. 
 We have investigated whether an underlying  scattered or direct H$\alpha$ broad component from the BLR is present,
which might be detected thanks to the lower reddening and larger strength for this line compared to H$\beta$.
Including a fourth component in the fit produces a successful
result. It has FWHM 2200$\pm$100 km s$^{-1}$ and a velocity
shift -35$\pm$50 km s$^{-1}$ relative to the narrow core of H$\alpha$. However, an H$\alpha$ component from
the BLR would be expected to be much broader and  this interpretation seems unlikely.
 
It is  on the other hand natural to expect that the outflowing gas emits also [NII] and H$\alpha$. We have thus assumed that the H$\alpha$ and [NII] lines consist, like [OIII],  of two kinematic components each. The following constraints must be applied in order to obtain physically meaningful fits: the FWHM of the narrow and broad components are the same for the H$\alpha$ and [NII]$\lambda\lambda$6548,6583 lines. 
The FWHM of the [NII] and H$\alpha$ lines is forced to be the same in km s$^{-1}$ as in [OIII]  for
the broad and narrow components.  The ratio between the two [NII]$\lambda\lambda$6548,6583 lines must be 1:3 for both kinematic components (Osterbrock 1989). The $z$ of the broad components on one hand, and the narrow components
on the other, are forced to be the same.  The fits imply that the narrow and broad components in the red lines are shifted by 430$\pm$100 km s$^{-1}$.

The main properties of the narrow and broad components are shown in Table 2. It can be seen that the line ratios are typical of AGNs in both cases.  Given the uncertainty on how both components are affected by underlying stellar absorption, this has been ignored in the current analysis, but this does
not affect our main conclusions. 

Therefore, SDSS J1430-00 contains a nuclear outflow with line FWHM$\sim$1600$\pm$200 km s$^{-1}$ and  shifted by 520$\pm$60 km s$^{-1}$ relative to the more quiescent gas. Independently of the outflow origin (stellar or AGN induced), the outflowing gas is ionized by the quasar and it must be located within the quasar ionization cone, consistent with the results by VM11b.

It is unknown whether the extended gas is clumpy or consists of diffuse gas that fills the slit, therefore the uncertainty on the IP is larger in this case. 
Taking this into account, the lines  have  FWHM   $\la$450 km s$^{-1}$. Unless slit effects play a role, the redshift  increases steeply outwards (Fig.~2), reaching a maximum shift in velocity relative to the narrow nuclear [OIII] core of +220$\pm$20 km s$^{-1}$ in the outer region of the extended nebula.

\section{HST images of SDSS type 2 quasars at 0.3$\la z \la$0.4.}

We present  a catalogue of HLA WFPC2 and ACS/WFC  images of 58 SDSS luminous type 2  AGNs (mostly quasars) at 0.3$\la z \la$0.4 (Table 3) from \cite{zak03} and \cite{rey08} catalogues. The images were obtained for the HST programme with identification 10880  and PI Henrique Schmitt. It was executed between August 2006 and November 2008.  In all cases
the total exposure time was 1200 seconds. These images provide rich information  about different issues related to type 2 quasars such as  their   morphological variety, their environment, the presence of merger/interaction signatures, etc. 

The catalogues by Zakamska et al. (2003) and Reyes et al. (2008) contain a total of
187 objects at 0.3$\le z \le$0.4. Of these, 109 have  [OIII] luminosity $\ga$10$^{8.3}$ L$_{\odot}$. This is the lower limit applied by the authors to ensure that the objects selected are quasars, rather than Seyferts (e.g. Hao et al. \citeyear{hao05}). Out of the 58 objects observed for the 10880 HST programme,
 42  qualify as type 2 quasars according to their [OIII] luminosity (Column [9], Table 3).

  In summary, the sub-sample observed with HST (contains 42 out of the 109 (38.5\%) SDSS type 2 quasars at 0.3$\la z \la$0.4. Since they span the same [OIII] luminosity range, we consider the HST sub-sample a good representation of all SDSS type 2 quasars in this $z$ range.
Moreover, the HST sub-sample  includes  16   very luminous Seyfert 2s in the same $z$ range.

The images of all objects are shown in Figs. 11 to 21. In general, they show a 20"$\times$20" field centered on each quasar.
In order not to miss interesting morphological features,  for a few exceptions a  somewhat larger size and/or a different spatial center of the field of view are shown.  Two images with different contrasts
 are presented for the same object when this is necessary to highlight interesting low and high surface brightness features.

The object names in the figures have been shortened. For instance SDSS J0025-10 is used instead of SDSS J002531.46-104022.2; SDSS J0210-01 replaces  SDSS J021059.66-011145.5, etc. In general, with a few exceptions applied to avoid leaving blank spaces, the quasars are organized according to increasing RA.   

 Table 3 shows information about the objects  and  the HST instrumental set up.  
We have classified visually the objects according to the morphological type of the galaxy (column [6]): Sp: disk with spiral arms;  Disk: disk with no obvious spiral arms; E: Elliptical; MS: multiple system (when 2 or more components are interacting and make the classification in the previous groups difficult). 
 It is found that most type 2 quasars are classified as ellipticals (30/42 or $\sim$71\%), while 7/42 (17\%) are classified as multiple systems (MS) 
 and only 4 are spirals or disks (9.5\%). One object is of dubious classification.  Thus, the majority of type 2 quasars are hosted by ellipticals and only a very small fraction are associated disk/spiral galaxies. Among the 16 luminous Seyferts, 8 are ellipticals (50\%), 7 (44\%) are disks or spirals
 and 1 (6\%) is a multiple system.  
  
Following the classification scheme proposed by  \cite{ram11}, we indicate in column [7] whether the objects show signatures of morphological disturbance. The nature of these features is indicated in column [8]: T: Tail; F: Fan; B: Bridge; S: Shell; D: Dust feature; 2N: Double Nucleus;  A: Amorphous halo; I: Irregular feature; IC: interacting companion.  It is found that a   significant fraction of type 2 quasars show signatures of morphological disturbance 25/42 ($\sim$59\%) which are suggestive of mergers/interactions.  The percentage is likely to be higher if deeper images were used, since these would allow to investigate  lower surface brightness level features.

\section{Discussion}

We have presented a compilation of HST images of 58 luminous SDSS type 2 AGNs   at 0.3$\la z \la$0.4 from which 42 are type 2 quasars.  These represent 38.5\% of the total number of type 2 quasars (109)
in \cite{rey08} and \cite{zak03}  catalogues in this $z$ range and span the same luminosity range. We consider this a good representation of the
total  0.3$\la z \la$0.4 sample. We find that the  majority of type 2 quasars are hosted by ellipticals (30/42 or 71\%) and only 9.5\% (4/42) are associated with spiral or disk galaxies.  But for one object of difficult classification, the rest (7/42 or 17\%) have very complex morphologies due to two or more interacting objects. 

This is consistent with studies of the host galaxy properties of radio loud and radio quiet type 1 quasars which show that their host galaxies are in general elliptical galaxies (e.g. Dunlop et al. \citeyear{dun03}). 

A significant fraction of type 2 quasars (25/42 or $\sim$59\%) show clear signatures of morphological disturbance which are in most cases clearly identified with merger/interaction processes. This fraction is likely to be higher if deeper images were obtained reaching lower surface brightness levels   (Bessiere et al. \citeyear{bess12}).

 In order to investigate whether mergers/interactions are responsible for  triggering  the nuclear activity in type 2 quasars, a careful comparison should be performed with a galaxy control sample of non active galaxies.  \cite{bess12} compare a complete sample of type 2 quasars at a mean redshift z = 0.35 with a sample of quiescent early type galaxies at the same median redshift. The control sample galaxies are also well matched in terms of morphology and luminosity to the quasar sample. 

They find that there is no significant difference in the proportion of galaxies that show evidence of tidal disruption between the two samples, with 75\% of the type 2 quasars and 68\% of the control sample galaxies showing evidence for morphological disturbance. However, they do find a significant difference in the surface brightnesses of the morphological features which are significantly brighter ($\sim$2 mag, r band) in the type 2 quasar sample. The authors conclude that this could highlight a difference in the type of systems that are merging in these two groups and that, if mergers are significant in triggering quasar activity, it is likely that factors such as the type of merger  (major or minor), orbital parameters of the encounter and gas content of the  merging galaxies, will also play a significant role in whether the encounter will lead to quasar activity. 

We have studied in detail the particular case of the radio quiet type 2 quasar SDSS J1430-00 at $z=$0.318 based on imaging and spectroscopic HST, VLT and SDSS data. This object offers an interesting scenario to investigate fundamental phenomena related to quasars and galaxy evolution in general: mergers/interactions, outflows (negative feedback) and star formation as well as the possible links among them.

SDSS J1430-00 is a member of a interacting system in which at least one of the galaxies involved must be rich in gas. 
It is associated with  a complex system of tidal tails  and  two confirmed nuclei
separated by  0.75" or 3.5 kpc in projection. The system is embedded in a large  low surface brightness amorphous halo, which shows rich substructure  and extends for $\sim$64$\times$69 kpc in the N-S and E-W directions respectively.  The longest tidal tail stretches towards the South for  at least 100 kpc from
the quasar. 

The morphology of the system  is quite similar to that often found in ULIRGs (e.g. Miralles-Caballero, Colina \& Arribas  \citeyear{mir11}).
We are observing the object in the final stage of a merger, before the nuclei coalesce. If the galactic interaction has triggered the nuclear activity in this object, this has occurred  before the coalescence (Ramos-Almeida et al. \citeyear{ram11}).   
 Following \cite{zau10} we estimate a dynamical time scale of $\sim$11 Myr for completion of the merger, assuming the projected distance between the nuclei represents their true separation, and that the nuclei are moving radially towards each other with a relative radial velocity of 300 km s$^{-1}$.

 A further similarity with the ULIRGs is that our spectral synthesis modeling results show evidence for a dominant YSP with age $<$80 Myr within the 3" (14 kpc) diameter aperture covered by the SDSS fibre; evidence for even younger YSP ($<$10 Myr) is provided by our results on the ionized gas which indicate local stellar photoionization in extranuclear regions at $r\sim$15 kpc in the Western tidal tail.   These results on the YSP are consistent with those obtained by Rodr\'\i guez Zaur\'\i n, Tadhunter  \& Gonz\'alez Delgado (\citeyear{zau10}, \citeyear{zau09})
who used similar spectral synthesis modeling to show that all the objects in their complete sample of 36 nearby ULIRGs show evidence for a YSP with luminosity-weighted ages $<$100~Myr. Such ages are also consistent with the $\sim$100 Myr timescale of the main merger-induced starburst predicted by hydrodynamical simulations of major, gas-rich mergers and the second episode when the nuclei are close to coalescence (e.g. Mihos \& Hernquist \citeyear{mih96}; Barnes \& Hernquist \citeyear{barn96};  di Matteo et al. \citeyear{dm07}) . 

Thus, SDSS J1430-00 shows  very similar properties (morphology, stellar populations) as found in ULIRGs. Is it actually a ULIRG? We have estimated the
infrared (IR) luminosity of the quasar using the
WISE flux measurements at 12 and 24 $\mu$m and the IRAS upper limits at 60 and 100 $\mu$m\footnote{Wise photometry:
$F_{12\mu m} = 11.76 \pm 0.03$ mJy; $F_{24\mu m} = 30.4\pm0.1$ mJy. IRAS upper limits: $F_{60 \mu m} <$ 100 mJy; $F_{100 \mu m} <$ 450 mJy.}.

 Three different scenarios have been considered. The prescription by \cite{san96} to
calculate the IR luminosities was used for the three
methods.

(i) Using the WISE flux measurements at 12 and 24 $\mu$m along with the IRAS upper limits at 60 and 100 $\mu$m, assuming that these are actual 60 and 100 $\mu$m fluxes. In this case L$_{IR}$=1.5$\times$10$^{12}$ L$_{\odot}$. This would place the object in the ULIRG regime. 
 
(ii) Using the WISE 12 and 24 $\mu$m fluxes and calculating the 60 and 100 $\mu$m fluxes by extrapolation, assuming a mid- to far-IR spectral index that is typical of radio galaxies at similar $z$ with no signs of star formation activity ($\alpha$=0.7) (Dicken et al. \citeyear{dick09}). In this case:
L$_{IR}$=0.8$\times$10$^{12}$ L$_{\odot}$, which is just below the ULIRG limit.

(iii) Using the WISE 12 and 24 $\mu$m fluxes and assuming that the 60 and 100 $\mu$m fluxes are the same as the WISE 24 $\mu$ flux ($\alpha$=0.0) --
i.e. an unusually blue SED. L$_{IR}$=0.6$\times$10$^{12}$ L$_{\odot}$.

The  last two cases are extremely conservative since they apply to AGN that have no signs of star formation, whereas the optical spectrum shows that
SDSS J1430-00 harbours very intense star formation activity. Therefore it would be expected to have redder mid- to far-IR colours (and steeper spectral index). 

We can confidently state that the IR luminosity of this source is in the range 6$\times$10$^{11}$ L$_{\odot} <$ L$_{IR} <$ 1.5$\times$10$^{12}$ L$_{\odot}$. If it
is not a ULIRG it is very close to this regime.

 The system presents a rich emission line spectrum. Using the relation between the H$\alpha$ luminosity L(H$\alpha$), the gas density  $n_e$ and mass of ionized gas
($M_{gas} \propto \frac{L(H\alpha)}{n_e}$, e.g. Holt et al. \citeyear{holt03}), we infer an ionized gas mass  $\sim$several$\times$10$^6$ M$_{\odot}$ in the nuclear
region.
In addition,
low surface brightness  extended line emission is detected at different spatial locations.  The maximum measured projected extension of the ionized gas  
is $r\sim$2.9" or $\sim$13 kpc from the quasar,  although very faint line emission  up to $r\sim$7" or 32 kpc might also be detected.
The characterization of the  spatial distribution and emission line ratios of the ionized gas reveals that both stellar and AGN photoionization are responsible for ionizing the gas, with varying relative contribution depending on the location. Extended ionized nebulae with hybrid ionizing mechanisms have been  found in other type 2 quasars (VM11a).

There is an ionized outflow in the central region ($r\la$few kpc) of SDSS J1430-00.  The outflowing gas is blueshifted by 520$\pm$50 km s$^{-1}$
relative to the  narrow [OIII] core, which we consider a good indicator of the systemic redshift (VM11b). It emits broad lines with FWHM=1600$\pm$200 km  s$^{-1}$. Thus, it not only emits very broad lines, but also shows a large blueshift relative to the systemic redshift.  Independently of the origin (whether starburst or AGN induced), the outflowing gas is photoionized by the quasar and is therefore within the quasar ionization cones.

Following the same procedure  and using the definitions as VM11b, we measure  $R2=\frac{F([OIII]_{broad})}{F([OIII]_{tot})}$=0.24$\pm$0.02 and $R=\frac{FWHM([OIII]_{broad})}{FWHM_{stars}}$=5.8$\pm$0.8 for the broad component. This object, therefore, follows the  correlation found by VM11b according to which, the more perturbed the outflowing gas is relative to the stellar motions, the lower  its relative contribution to the total line flux  is.

We define $R3$  as the ratio between the H$\alpha$ flux from the outflowing gas and the total line flux. For SDSS J1430-00, $R3=\frac{F(H\alpha_{broad})}{F(H\alpha_{tot})}$=0.12$\pm$0.03.  Given the dependence of the ionized gas mass with density and line luminosity ($M_{gas} \propto \frac{L(H\alpha)}{n_H}$), this  $R3$  value implies that,  for similar densities and line reddening, only $\sim$12\%  of the nuclear ionized gas ($r\la$few kpc), i.e. $\sim$several$\times$10$^{5}$ M$_{\odot}$ participates in the outflow. This is likely to be an upper limit since the outflow can produce a shock induced density enhancement  (e.g. Villar-Mart\'\i n et al. \citeyear{vil99}). As already argued in VM11b, to characterize the full impact of the outflow phenomenon in this and other type 2 quasars, it is essential to investigate the outflow effects in other gaseous phases such as molecular and neutral. For an illustrative comparison, the masses involved in
neutral and molecular outflows in Seyfert and  starburst ultraluminous
infrared galaxies (ULIRGs),  which are constrained to spatial scales of $<$few kpc, are often $>$10$^8$ M$_{\odot}$ (e.g. Rupke, Veilleux \& Sanders \citeyear{rup05}, Sturm et al.  \citeyear{sturm11}).

 We propose the following scenario to explain the general properties of SDSS J1430-00 in the context of theoretical and observational studies  of galaxy mergers/interactions and their role in the triggering of the nuclear and star formation activities in the most luminous AGNs:  A first pericenter passage between two companion galaxies has already occurred. At least one of them must have been gas rich. This or both galaxies have been  heavily disrupted in the process. Tidal tails   have developed as a consequence. The removal of gas from the central regions has not prevented the accumulation of enough gas in the quasar nuclear region  to enhance the star formation activity and the triggering of the quasar soon before the coalescence of the two nuclei.   Extranuclear regions at $r\sim$15 kpc contribute as well to the star formation activity (e.g. di Matteo et al. \citeyear{dm07}, di Matteo, Springel \& Hernquist  \citeyear{dm05}, Barnes \& Hernquist \citeyear{barn96}).  

The nuclear starburst started less than 80 Myr ago.
Since the  quasar lifetime is likely to be in the range $\sim$1-100 Myr (e.g. Haiman \& Hui \citeyear{hai01}, Hopkins et al. \citeyear{hop05}) the fact that they coexist implies that the starburst and the quasar were triggered within a time interval of $\la$several$\times$10$^7$ yr. The galaxy merger has possibly triggered both (Bessiere et al. \citeyear{bess12}, Ramos-Almeida et al. \citeyear{ram11}).  

At certain point a wind was generated  in the nucleus, probably associated with one or both events, i.e. the starburst and the triggering of the quasar (di Matteo, Springel \& Hernquist  \citeyear{dm05}).   $\la$several$\times$10$^5$ M$_{\odot}$ of ionized gas mass are involved, which is unlikely to have a significant  impact on the future evolution of both the quasar and the star formation activities. The ionized outflow, on the other hand, might hint the existence  of a more powerful wind involving a large mass of molecular and neutral gas. What  is clear is that  such wind, if existing, has not been powerful enough or it has been triggered too late to prevent the most recent burst of star formation.

\section{Conclusions}

 We present a compilation of HST images of 58 luminous SDSS type 2 AGNs  at 0.3$\la z \la$0.4 among which 42 are type 2 quasars. These represent $\sim$38\% of the total number of type 2 quasars
in \cite{rey08} and \cite{zak03}  SDSS catalogues in this $z$ range. Since they span the same luminosity range, we consider the HST sub-sample  a good representation of all SDSS type 2 quasars at 0.3$\la z \la$0.4. We find that the majority of the host galaxies are ellipticals (30/42 or 71\%). 7/42 (17\%) have very complex morphologies due to two or more interacting objects while only 4 (9.5\%) are associated with spiral or disk galaxies. This is consistent with studies of radio loud and radio quiet type 1 quasars which show that their host galaxies are in general ellipticals.

A significant fraction of type 2 quasars (25/42 or $\sim$59\%, which is a lower limit) show clear signatures of morphological disturbance which are in most cases clearly identified with merger/interaction processes. We discuss this in the context of
 related works on  type 2 quasars and powerful radio galaxies.

We have studied in detail the particular case of the radio quiet type 2 quasar SDSS J1430-00 at $z=$0.318 based on VLT, HST and SDSS imaging and spectroscopic data.  It is a member of a interacting system in which at least one of the galaxies involved must be rich in gas. It shows a complex morphology, similar to that frequently found in ULIRGs such as a double nuclei and several tidal tails.  Indeed, the IR luminosity of this quasar is expected to be in the ULIRG regime or close to it. Our spectral synthesis modeling results show evidence for a dominant YSP with age $<$80 Myr within the central 3Ó ($\sim$14 kpc) ; evidence for even younger YSP ($<$10 Myr) is also found   in extranuclear regions at$\sim$15 kpc.  It is possible that both the nuclear activity and the starburst are a consequence  of the galactic interaction. In such case, both have been triggered before the nuclei coalesce. Hydrodinamical simulations show that this is feasible.

As frequently found in optically selected type 2  quasars at similar $z$, a nuclear ionized outflow has been identified in SDSS J1430-00. The outflowing gas is blueshifted by 520$\pm$50 km s$^{-1}$ relative the systemic redshift  and  emits broad lines with FWHM=1600$\pm$200 km s$^{-1}$. Independently of the origin, the outflowing  gas has been ionized by the quasar.  It is expected that  $\la$several$\times$10$^5$  M$_{\odot}$ at most participate in the ionized outflow, well below the $>$10$^8$ M$_{\odot}$ involved in neutral and molecular outflows of some starburst and AGN ULIRGs. The ionized flow is unlikely to have a significant impact on the system's evolution.  It is essential to investigate the presence of neutral and molecular outflows in this and other type 2 quasars to evaluate the full impact as a feedback mechanism in the surrounding medium and, ultimately, the system's evolution.

 Ionized gas is detected up  $r\sim$13 kpc from the quasar, although gas at   $r\sim$32 kpc has been possibly detected. The ionizing mechanism,
AGN vs. stellar photoionization, varies depending on the spatial location.  Extended ionized nebulae with hybrid ionizing mechanisms have been  found in other type 2 quasars.

We discuss the global properties of this quasar in the context of theoretical and observational studies  of galaxy mergers/interactions and their role in the triggering of the nuclear and star formation activities in the most luminous AGNs.

\section*{Acknowledgments}
This work has been funded with support from the Spanish Ministerio de Ciencia e Innovaci\'on through the grants 
AYA2007-64712 and AYA2010-15081.   Thanks to  the staff at Paranal Observatory for their support during the observations.

\newpage
 \begin{table*}
\small
\begin{tabular}{lllllllll}
\hline
Object & z & L[OIII] &    Instrument/Detector &  Filter  & Morph. Type & Disturbed? &  Features  & Quasar? \\ 
~[1]  &  [2]  &  [3]  &  [4]  &  [5]  &  [6]  &  [7]  & [8] & [9]  \\ \hline
SDSS J002531.46-104022.2 	& 0.303 &	8.73	&    ACS/WFC &    F775W & MS	&   Yes & T,2N  & Yes \\   
SDSS J002852.87-001433.6 & 0.310 & 8.43 & WFPC2	&  F814W  & E  & Yes  &  T    & Yes\\ 
SDSS J005515.82-004648.6 & 0.345 & 8.15 &    WFPC2 	& F814W &  E & U  &  & No \\ 
SDSS J011429.61+000036.7  &  0.389 & 8.66  &   ACS/WFC  &   F775W   & E & U &  & Yes \\  
SDSS J011522.19+001518.5  & 0.390 & 8.14   &    WFPC2 	& F814W  & E  &  Yes  & A,S? & No \\  
SDSS J014237.49+144117.9 &  0.389 & 8.76 &     ACS/WFC    &  F775W   & E &  U  & & Yes \\    
SDSS J015911.66+143922.5  &  0.319 & 8.56  &     ACS/WFC  &  F775W  & E & U &   &Yes \\  
SDSS J020234.56-093921.9  & 0.302   & 8.39  &    WFPC2 	& F814W  &  Sp & U &  ???   & Yes \\  
SDSS J021059.66-011145.5 & 0.384 & 8.10 &     WFPC2 &  F814W  & Sp & U &     & No \\  
SDSS J021758.18-001302.7   &  0.344 & 8.55  &      ACS/WFC &  F775W  &  E & Yes & T,D  & Yes \\
SDSS J021834.42-004610.3  & 0.372 & 8.85   &     ACS/WFC &   F775W	&  MS &  Yes & T,IC    & Yes \\ 
SDSS J022701.23+010712.3  & 0.363 & 8.90 &      ACS/WFC &   F775W   & E &  Yes & D,S,T,A  & Yes  \\  
SDSS J023411.77-074538.4 & 0.310 & 8.77 &   ACS/WFC  &    F775W	  &E &  U  & & Yes  \\  
SDSS J031946.03-001629.1 & 0.393 & 8.24 &    WFPC2 & 	F814W &  Disk & U? &  & No \\  
SDSS J031927.22+000014.5  &  0.385 & 8.06   &    WFPC2 	& F814W  & E & Yes &  F  & No \\ 
SDSS J032029.78+003153.5  &  0.384 & 8.52 &    ACS/WFC & F775W   &  E & Yes & D,2N?    & Yes \\    
SDSS J032533.33-003216.5  &  0.352 & 9.06 &    WFPC2 & 	 F814W  & E & Yes & T    & Yes  \\  
SDSS J033310.10+000849.1 &  0.327 & 8.13  &    WFPC2 & 	 F814W & MS & Yes & T,IC    & No  \\   
SDSS J034215.08+001010.6   & 0.348  & 9.08   &    WFPC2 	& F814W   & E &  U &   & Yes  \\
SDSS J040152.38-053228.7  & 0.320 & 8.96 &    WFPC2 	& F814W & E & U &   & Yes   \\ 
SDSS J074811.44+395238.0 &  0.372 & 8.19 &    WFPC2 	& F814W  & E & Yes & T   & No  \\  
SDSS J081125.81+073235.3 & 0.350  & 8.88  &    WFPC2 	& F814W & E & Yes & IC?, I   & Yes \\ 
SDSS J081330.42+320506.0  &  0.398 & 8.83 &   ACS/WFC &   F775W	&  E? & Yes & T,D,I,A   & Yes  \\    
SDSS J082449.27+370355.7 &  0.305 & 8.28 &    WFPC2	& F814W   & E &  U &    & Yes \\   
SDSS J082527.50+202543.4  & 0.336 & 8.88 &   	WFPC2	& F814W  & E & U &    & Yes  \\  
SDSS J083028.14+202015.7 & 0.344 & 8.91 &  	WFPC2	& F814W    & E &  U &   & Yes \\
SDSS J084041.08+383819.8  &  0.313 & 8.47     &    	WFPC2	& F814W  & Sp & Yes & T  & Yes   \\  
SDSS J084309.86+294404.7 &  0.397  & 9.34    &   	WFPC2	& F814W   & E & U &    & Yes \\ 
SDSS J084856.58+013647.8  &   0.350 & 8.46 &      ACS/WFC&   F775W & E	& Yes & S   & Yes  \\ 
SDSS J084943.82+015058.2  & 0.376  & 8.06 &    	WFPC2	& F814W &   Disk  & U &      & No \\
SDSS J090307.84+021152.2  & 0.329 & 8.42 &  	WFPC2	& F814W  & MS &  Yes &  T,2N,D   & Yes   \\
SDSS J090414.10-002144.9 & 0.353 & 8.93 &       ACS/WFC &   F775W	& E & Yes & T,F    & Yes  \\  
SDSS J090801.32+434722.6  & 0.363 & 8.31 &    	WFPC2	& F814W    &  E &Yes & T,I  & Yes \\
SDSS J092318.06+010144.8  &    0.386 & 8.94 &   	WFPC2	& F814W & E & Yes & T  & Yes   \\
SDSS J092356.44+012002.1 & 0.380 & 8.59 &   	WFPC2	& F814W   & E & Yes & T   & Yes \\ 
SDSS J094209.00+570019.7 & 0.350 & 8.31 &     	WFPC2	& F814W& E & Yes & T,F,D     & Yes  \\ 
SDSS J094350.92+610255.9 & 0.341 & 8.46 	 &     	WFPC2	& F814W    &  E & U &   & Yes \\
SDSS J095629.06+573508.9 & 0.361 & 8.38 &     	WFPC2	& F814W  & E &  U &    & Yes  \\
SDSS J100329.86+511630.7 & 0.324 & 8.11  &     	WFPC2	& F814W  & E & Yes & T,F    & No\\
SDSS J103639.39+640924.7  &  0.398 & 8.42 &     	WFPC2	& F814W   & E & U &   & Yes \\
SDSS J112907.09+575605.4 &  0.313  & 9.38  &      	WFPC2	& F814W  & E &  Yes & I    & Yes \\ 
SDSS J113710.78+573158.7 &  0.395 & 9.61 &   	WFPC2	& F814W  & E & Yes & T   & Yes  \\
SDSS J133735.01-012815.7  & 0.329 & 8.72 &    	WFPC2	& F814W  &  E & Yes & T?,2N,I   & Yes  \\
SDSS J140740.06+021748.3  &  0.309 & 8.90 &     	WFPC2	& F814W  & Disk &  U &  &Yes   \\
SDSS J143027.66-0056149 & 0.318 & 8.44 &    	WFPC2	& F814W   &  MS & Yes & T,2N,A,IC?   & Yes  \\
SDSS J144711.29+021136.2  & 0.386 & 8.45 &    WFPC2	& F814W   & MS &  Yes &  T, 2N?  & Yes \\
SDSS J150117.96+545518.3 & 0.338 & 9.06 &   	WFPC2	& F814W   &  E & Yes & B,S,D  & Yes \\
SDSS J154133.19+521200.1 & 0.311 & 8.25 &  	WFPC2	& F814W   & Sp & Yes & S   & No \\
SDSS J154337.81-004420.0 & 0.311 & 8.40 &    	WFPC2	& F814W    & MS & Yes & T,F,A   & Yes \\
SDSS J154613.27-000513.5 &  0.383 & 8.18 &     	WFPC2	& F814W    &  E & Yes &  T & No \\
SDSS J172419.89+551058.8 &  0.365 & 8.00 &     	WFPC2	& F814W  &  E & U &    &  No    \\
SDSS J172603.09+602115.7  & 0.333 & 8.57 &     	WFPC2	& F814W & E & U &      & Yes  \\
SDSS J173938.64+544208.6  & 0.384 & 8.42 &    	WFPC2	& F814W &  Disk &U &      & Yes  \\
SDSS J214415.61+125503.0  & 0.390 & 8.14 &   	WFPC2	& F814W  & E & Yes & I   & No\\
SDSS J215731.40+003757.1  & 0.390 & 8.39 &   	WFPC2	& F814W  & MS & Yes & D,IC    & Yes  \\
SDSS J223959.04+005138.3 &  0.384 & 8.15 &    	WFPC2	& F814W  & Disk &  U &   & No  \\
SDSS J231755.35+145349.4  & 0.311 & 8.10 &    	WFPC2	& F814W & Sp & U &    & No  \\
SDSS J231845.12-002951.4   &  0.397 & 8.00 &      	WFPC2	& F814W & Sp & U &   &  No  \\ 
 \end{tabular}
\caption{List of objects observed for the HST programme 10880.  The [OIII] luminosity [3] is given in log and relative to the solar luminosity (Zakamska et al. 2003, Reyes et al. 2008). According to the selection criteria of these authors, objects with L[OIII]$\ga$8.3  are type 2 quasars (column [9]). Objects with lower [OIII] luminosities fall in the regime of luminous Seyfert 2s (indicated with ``No" in the last column). We indicate in column [6] the morphological type: E (elliptical), Sp (spiral), Disk (disk galaxy with no obvious spiral arms), MS (multiple system: when 2 or more components are interacting and make the classification in the previous groups difficult). In column [7] we specify whether the object shows morphological peculiarities (``Yes") or look
undisturbed (``U"). The nature of these peculiar features is shown in [8] following the same classification as Ramos-Almeida et al. (2001). T: Tidal tail; F: Fan; B: Bridge; S: Shell; D: Dust feature; 2N: Double Nucleus;  A: Amorphous Halo; I: Irregular feature. A ``?" indicates uncertain classification or identification.}
\end{table*}

 \begin{figure*}
\includegraphics[width=11cm]{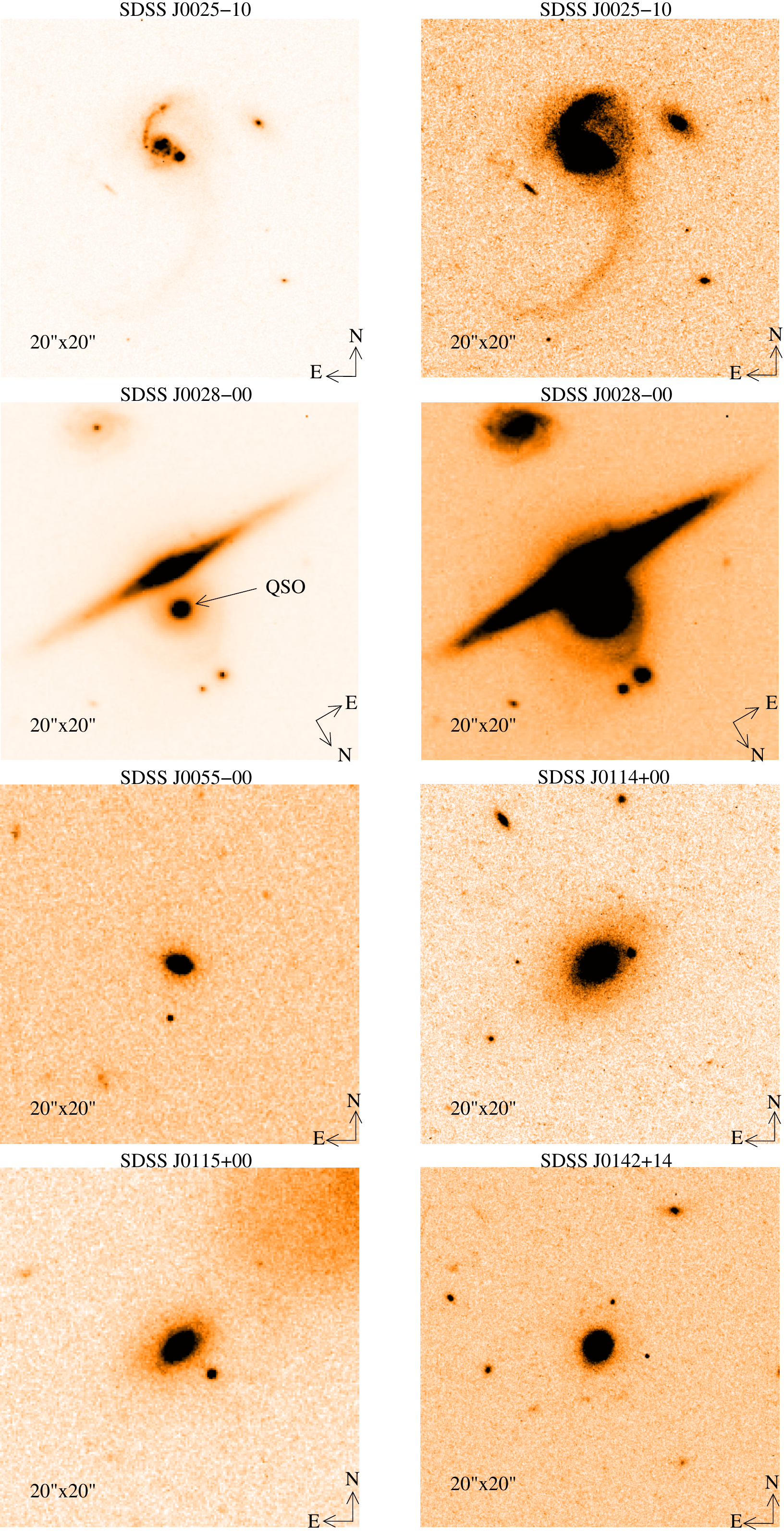}
\caption{HST images of type 2 quasars and luminous Sy2 galaxies at 0.3$\la z \la$0.4. Two images with different contrasts
 are presented for the same quasar when this is necessary to highlight interesting low and high surface brightness features. In general, a 20"$\times$20" field centered on the quasar is shown in each image. For a few exceptions,
 in order not to miss  interesting features a larger field of view or a different spatial center are chosen.}
\end{figure*}

\begin{figure*}
\includegraphics[width=11cm]{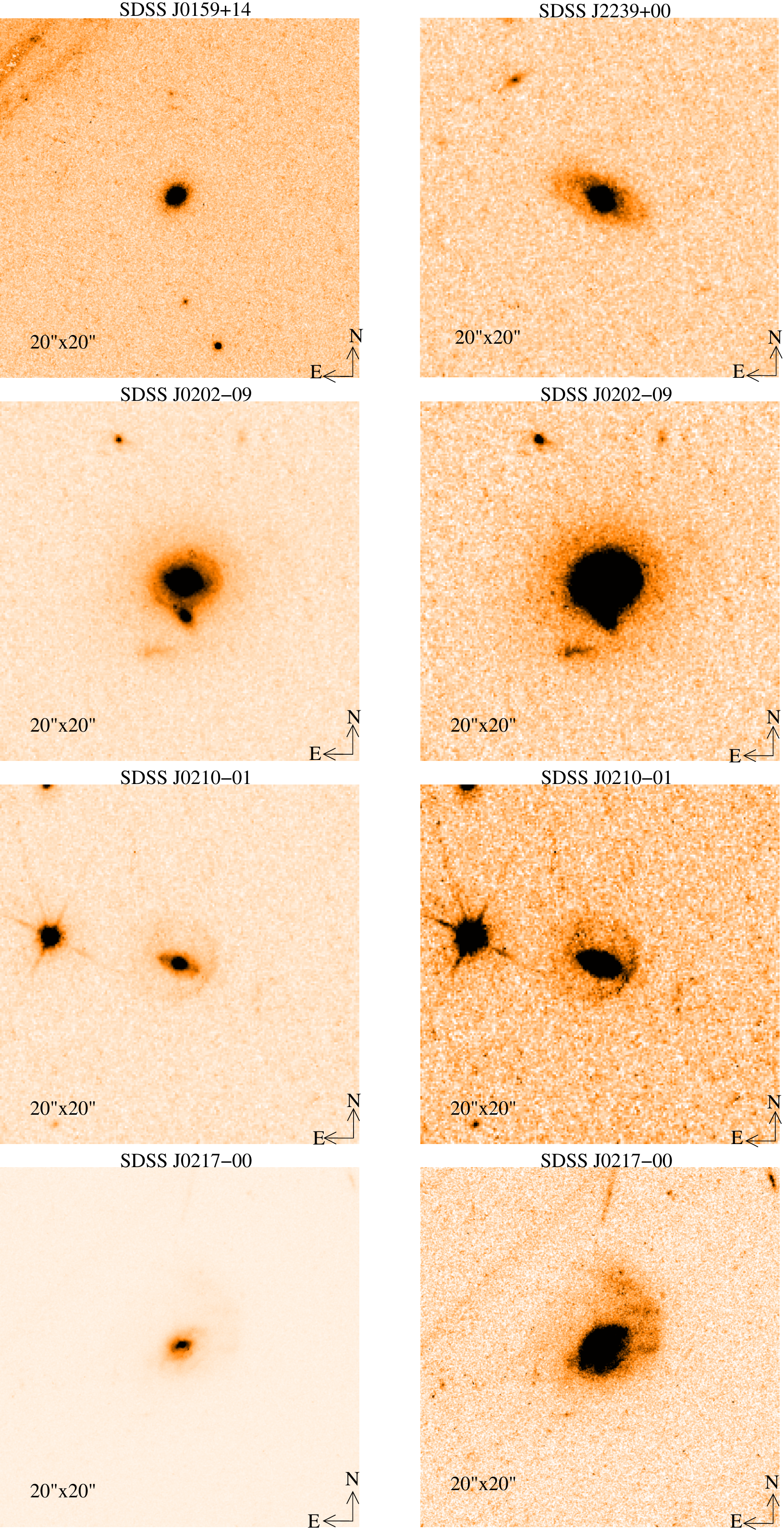}
\caption{HST images of type 2 quasars and luminous Sy2 galaxies (cont.)}
\end{figure*}

\begin{figure*}
\includegraphics[width=11cm]{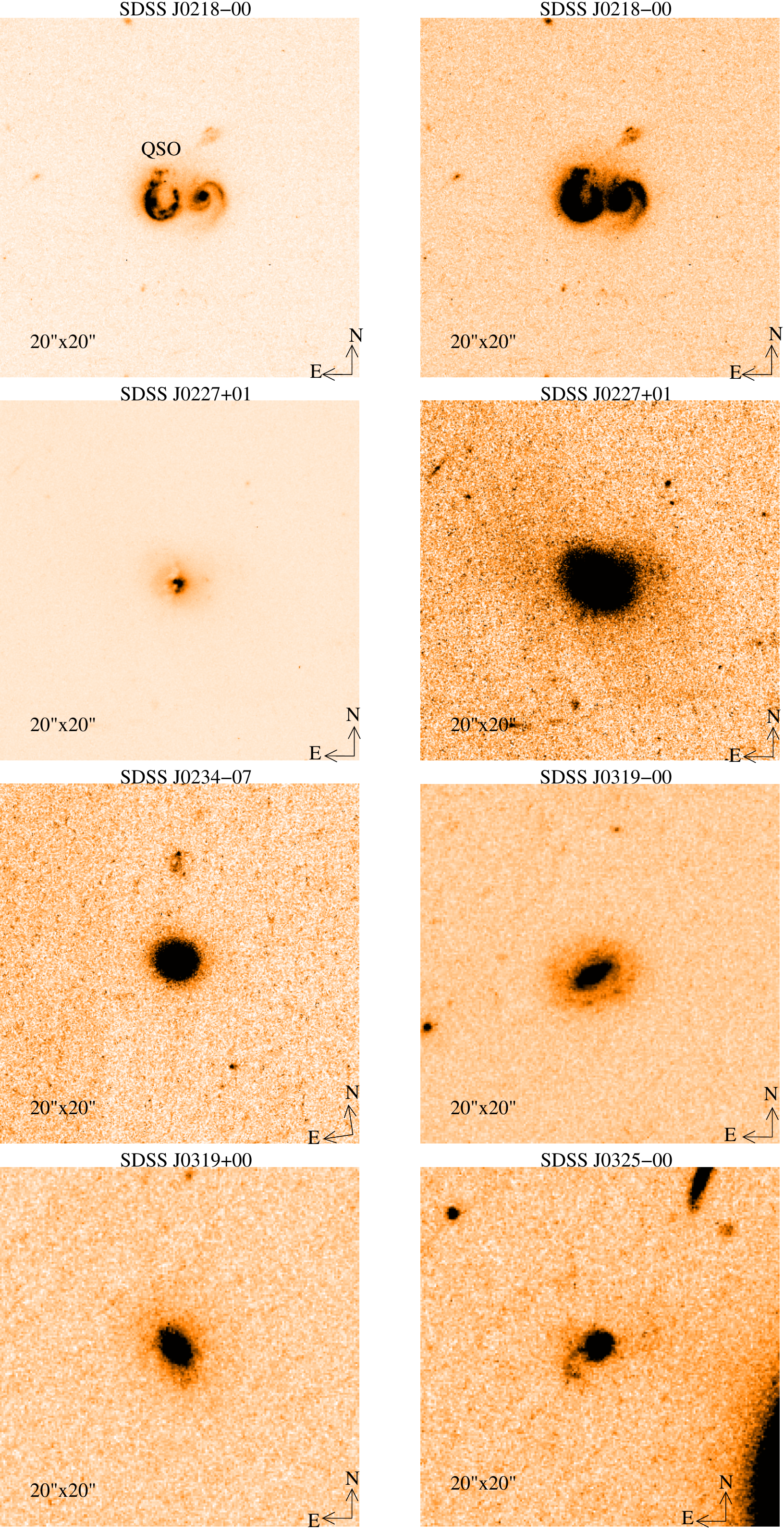}
\caption{HST images of type 2 quasars and luminous Sy2 galaxies (cont.)}
\end{figure*}

\begin{figure*}
\includegraphics[width=11cm]{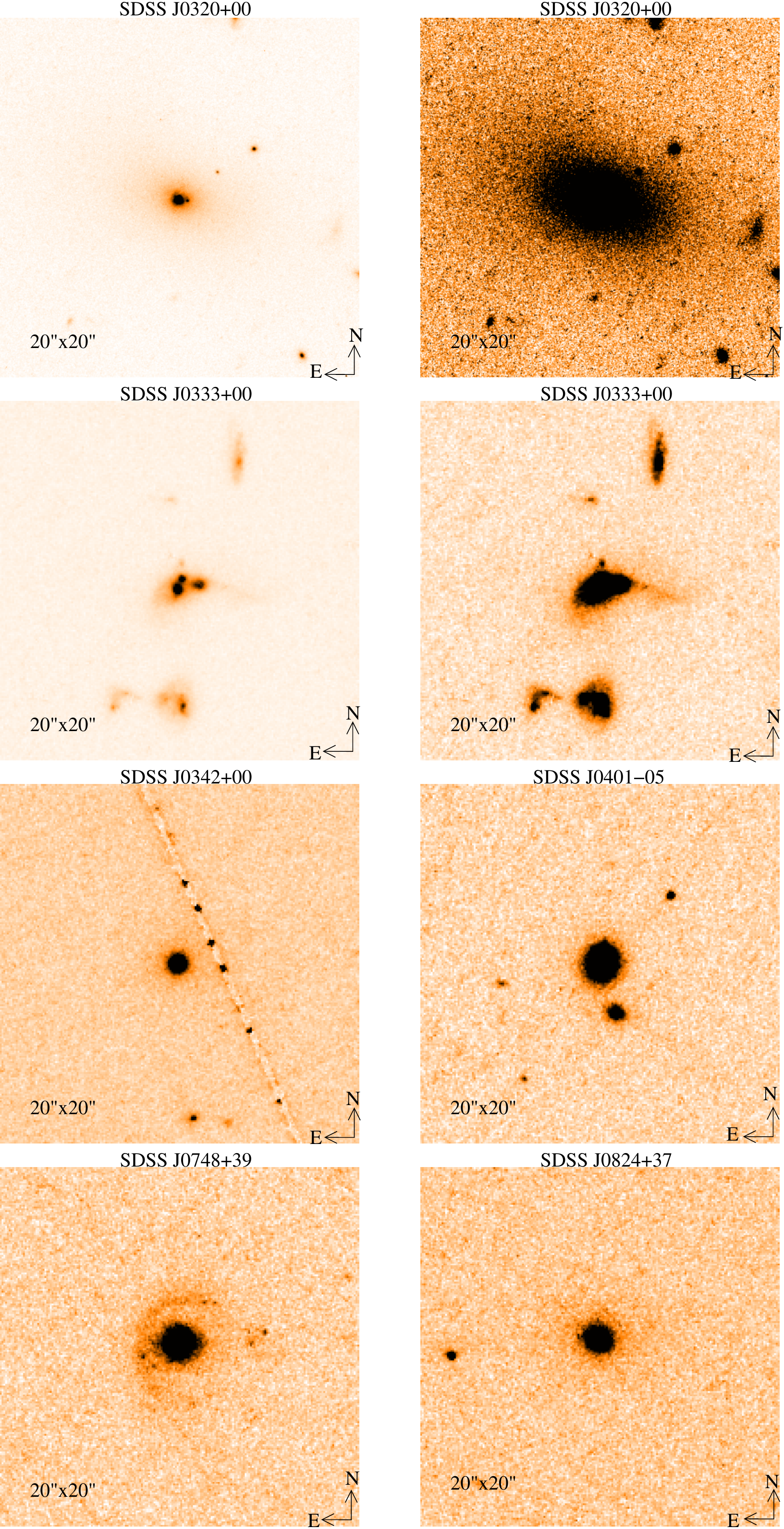}
\caption{HST images of type 2 quasars and luminous Sy2 galaxies (cont.)}
\end{figure*}

\begin{figure*}
\includegraphics[width=11cm]{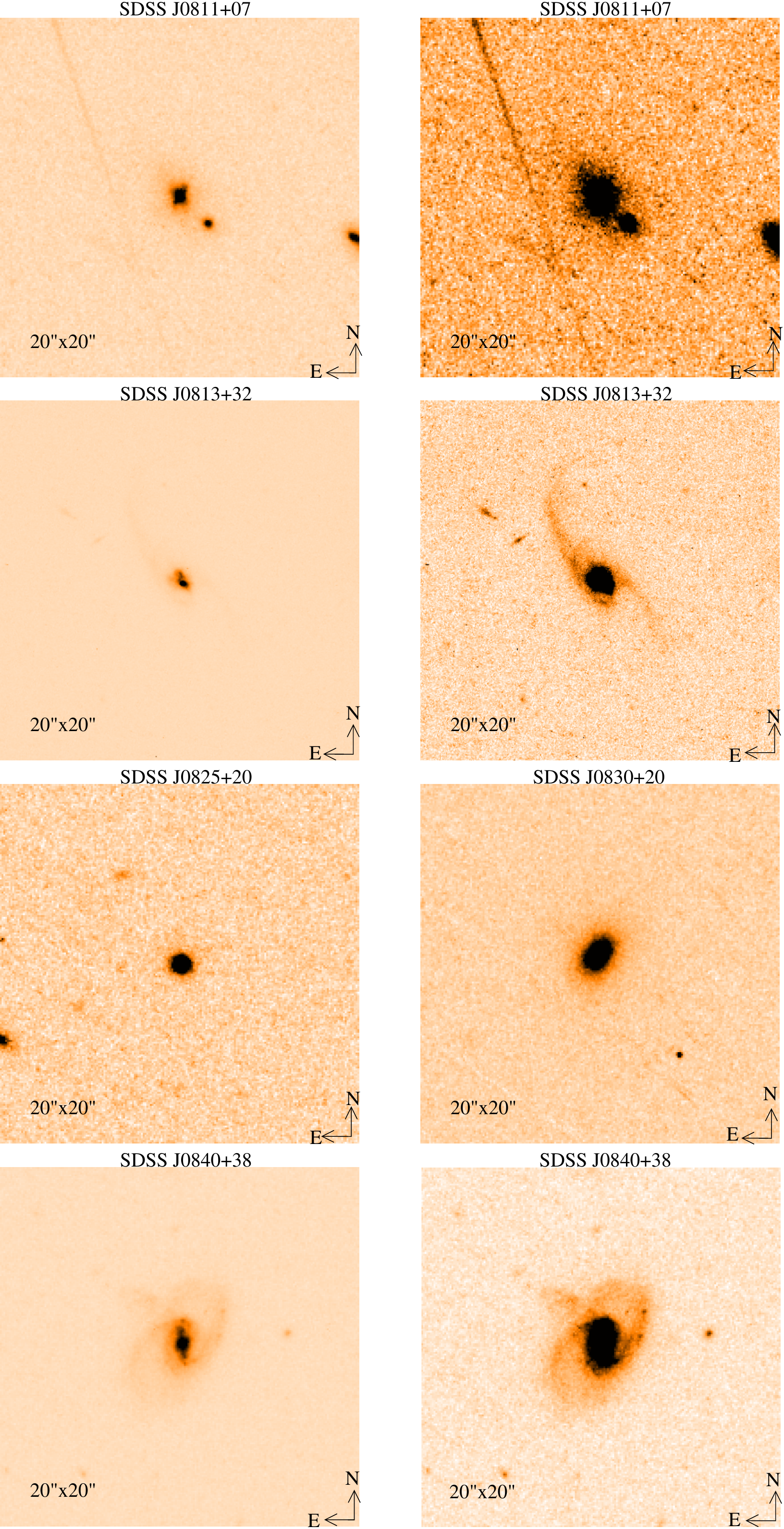}
\caption{HST images of type 2 quasars and luminous Sy2 galaxies(cont.)}
\end{figure*}

\begin{figure*}
\includegraphics[width=11cm]{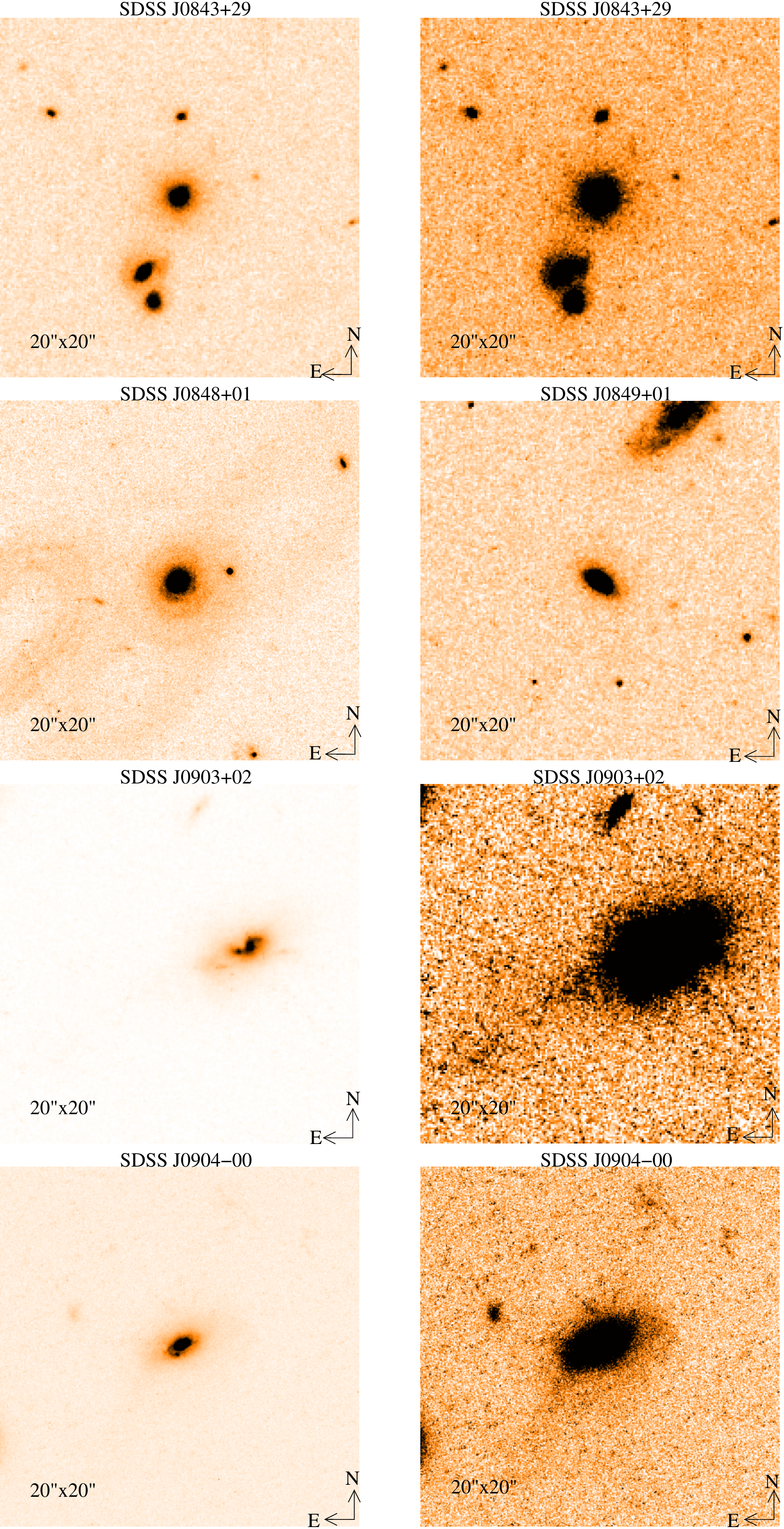}

\caption{HST images of type 2 quasars and luminous Sy2 galaxies (cont.)}
\end{figure*}

\begin{figure*}
\includegraphics[width=11cm]{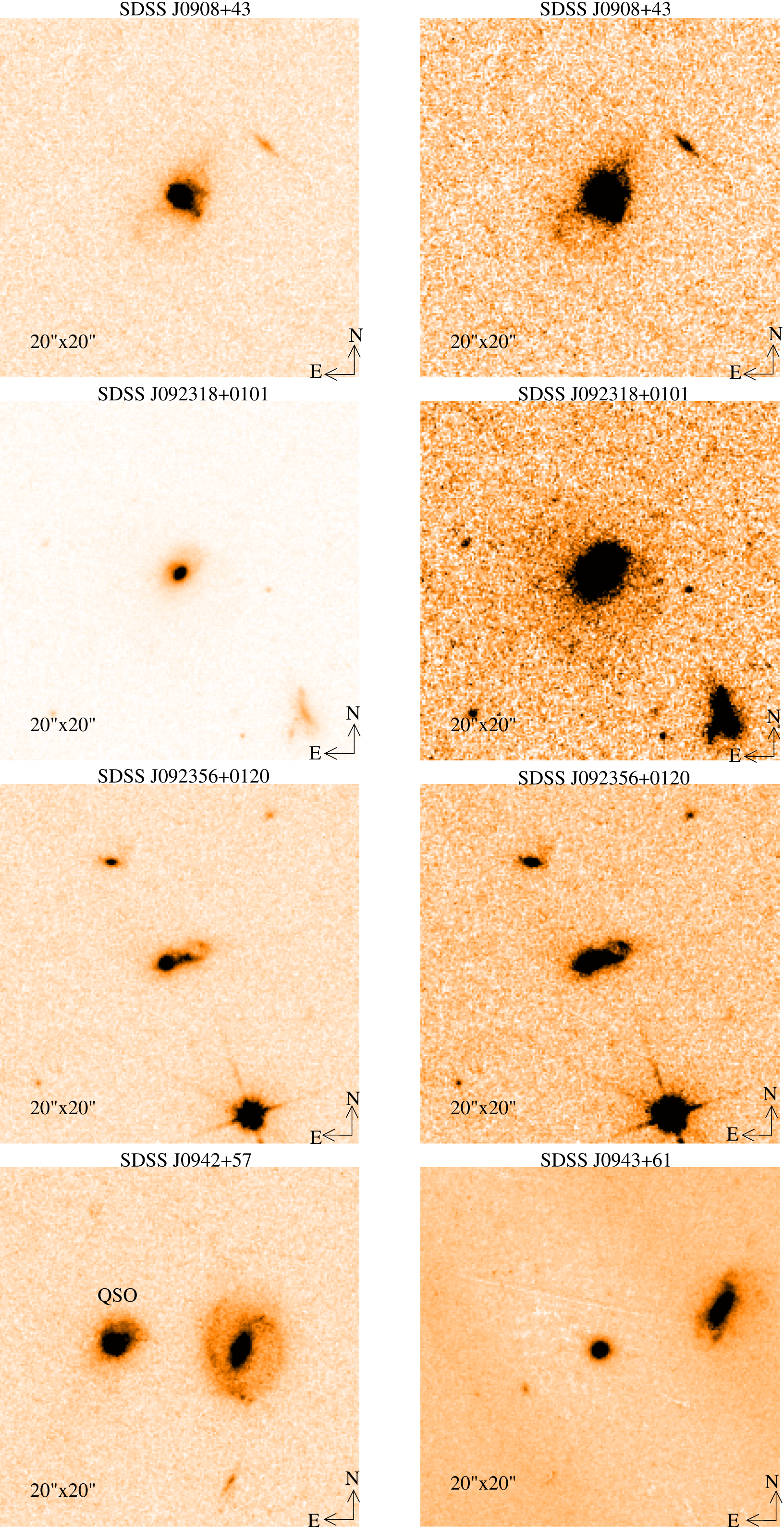}
\caption{HST images of type 2 quasars and luminous Sy2 galaxies (cont.)}
\end{figure*}

\begin{figure*}
\includegraphics[width=11cm]{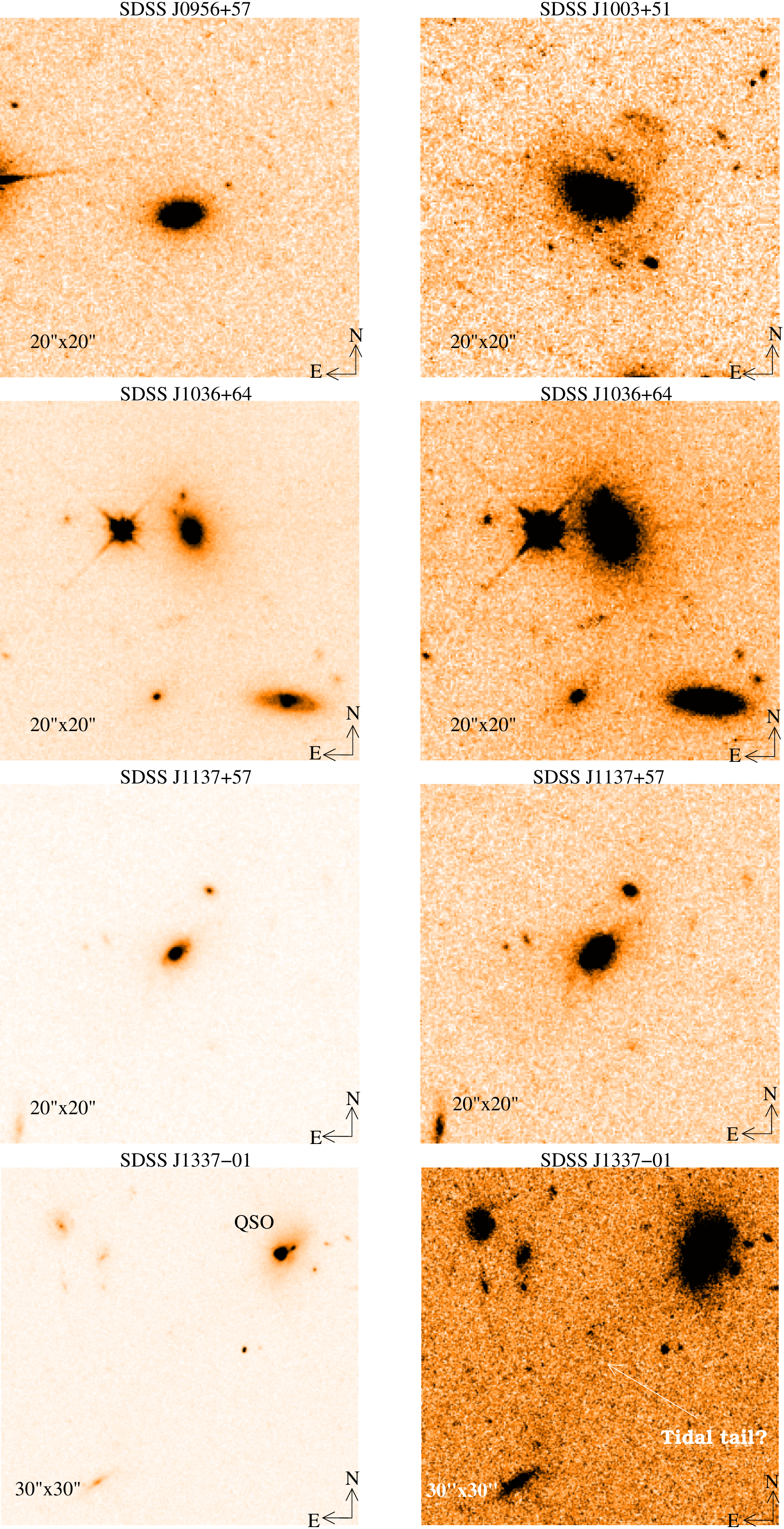}
\caption{HST images of type 2 quasars and luminous Sy2 galaxies (cont.)}
\end{figure*}

\begin{figure*}
\includegraphics[width=11cm]{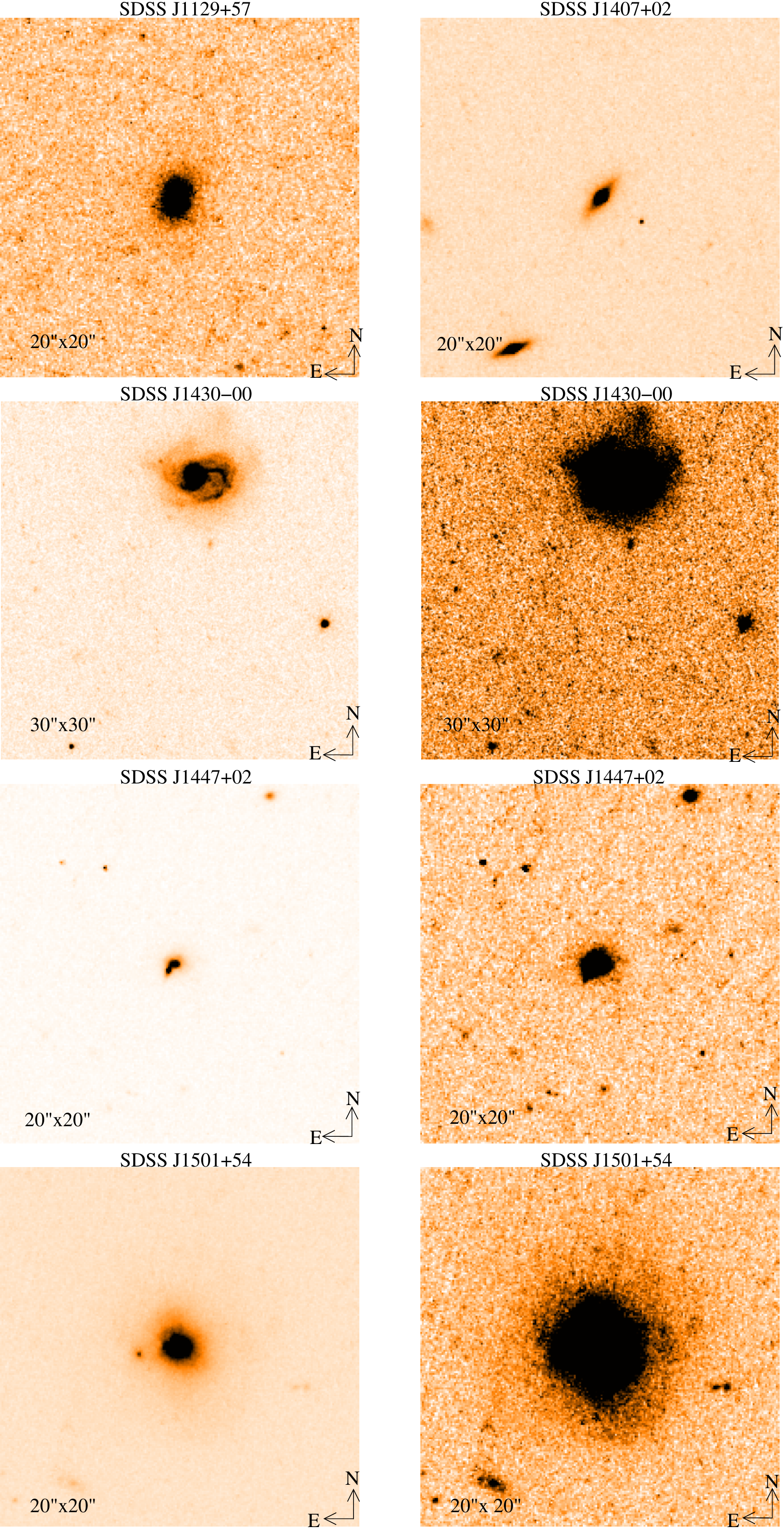}
\caption{HST images of type 2 quasars and luminous Sy2 galaxies (cont.)}
\end{figure*}

\begin{figure*}
\includegraphics[width=11cm]{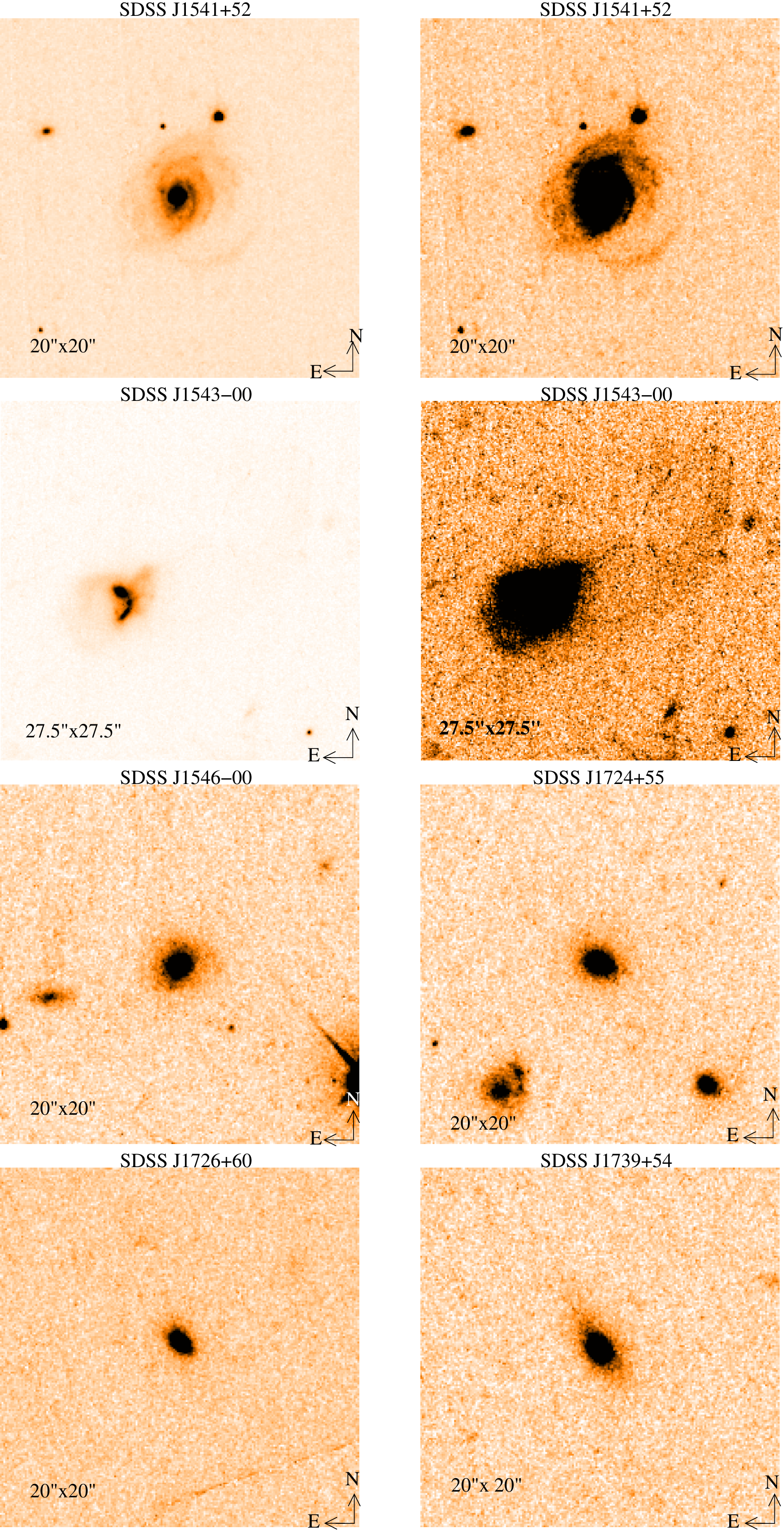}
\caption{HST images of type 2 quasars and luminous Sy2 galaxies (cont.)}
\end{figure*}

\begin{figure*}
\includegraphics[width=11cm]{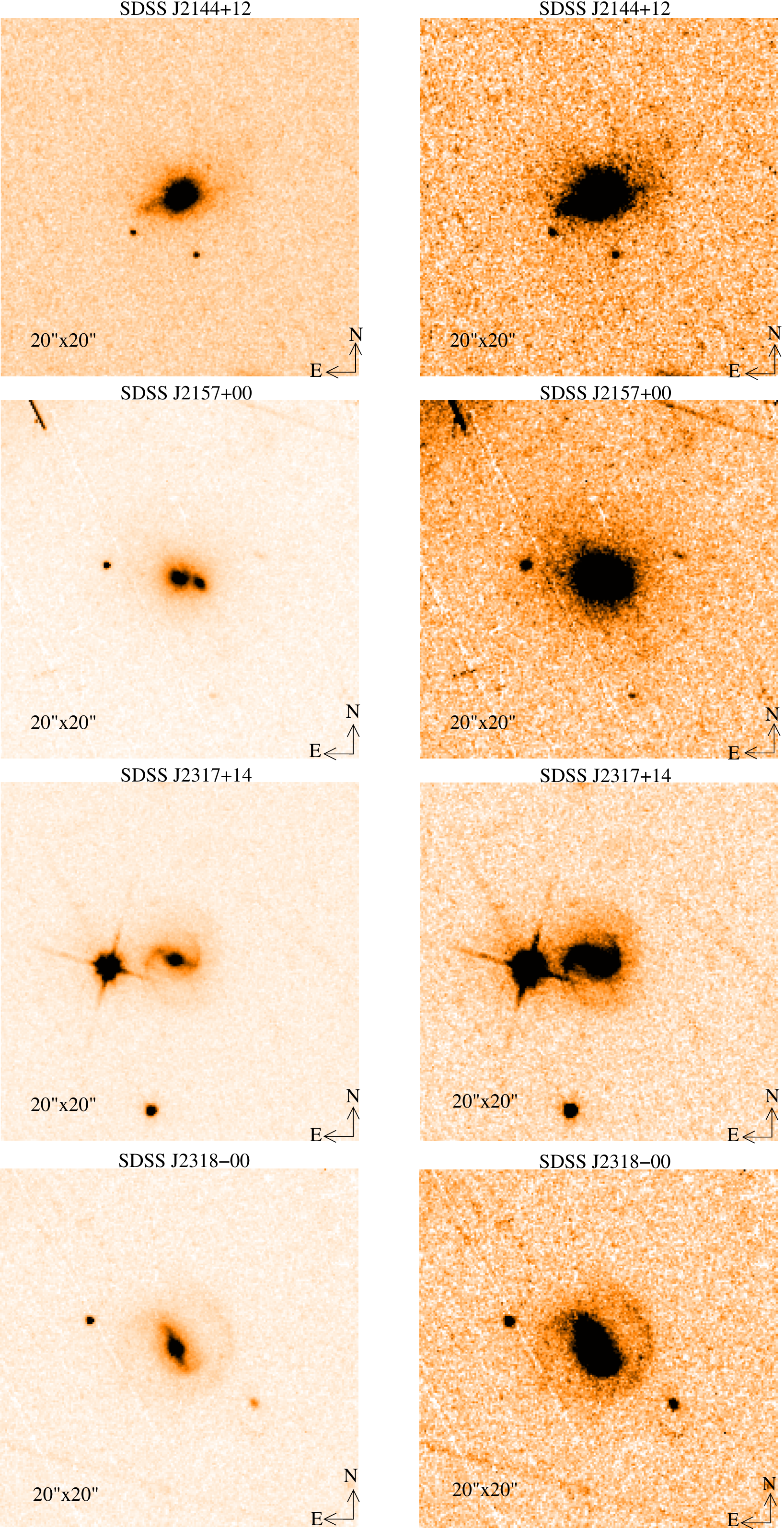}
\caption{HST images of type 2 quasars and luminous Sy2 galaxies (cont.)}
\end{figure*}

\end{document}